\documentclass[12pt,a4paper]{article}
\pdfoutput=1
\usepackage{jcappub}
\usepackage{graphicx,amssymb,amsmath,xcolor}
\usepackage[normalem]{ulem}
\graphicspath{{figs/}}

\newcommand{\be}{\begin{equation}}
\newcommand{\ee}{\end{equation}}

\def\lsim{\raise0.3ex\hbox{$\;<$\kern-0.75em\raise-1.1ex\hbox{$\sim\;$}}}
\def\gsim{\raise0.3ex\hbox{$\;>$\kern-0.75em\raise-1.1ex\hbox{$\sim\;$}}}

\begin{document}

\title{%
	  UHECR mass composition at highest energies from anisotropy of their arrival directions
  }

\author[a,b]{M.Yu.~Kuznetsov,}
\author[a,b]{P.G.~Tinyakov}

\affiliation[a]{Service de Physique Theorique, Universite Libre de Bruxelles, Boulevard du Triomphe CP225, 1050 Brussels, Belgium}
\affiliation[b]{Institute for Nuclear Research of the Russian Academy of Sciences, 60th October Anniversary Prospect 7a, 117312 Moscow, Russia}

\abstract{
We propose a new method for the estimation of ultra-high energy cosmic ray (UHECR) mass composition from a distribution of their arrival directions. The method employs a test statistic (TS) based on a characteristic deflection of UHECR events with respect to the distribution of luminous matter in the local Universe. Making realistic simulations of the mock UHECR sets, we show that this TS is robust to the presence of galactic and non-extreme extra-galactic magnetic fields and sensitive to the mass composition of events in a set. This allows one to constrain the UHECR mass composition by comparing the TS distribution of a composition model in question with the data TS, and to discriminate between different composition models. While the statistical power of the method depends somewhat on the MF parameters, this dependence decreases with the growth of statistics. The method shows good performance even at GZK energies where the estimation of UHCER mass composition with traditional methods is complicated by a low statistics. 
}
\emailAdd{mkuzn@inr.ac.ru}

\maketitle

\section{Introduction}
\label{sect:Introduction}

Despite the experimental progress in detection of ultra-high energy cosmic
rays (UHECR) and growing quality and quantity of data, our understanding of
this  phenomenon is hampered by three coupled unsolved problems:
UHECR sources, nature of UHECR particles and cosmic magnetic fields.

Identification of sources of UHECR from their sky distribution is not
straightforward.  While the arrival directions of incident particles are
reconstructed quite accurately --- with the precision of $1-1.5^\circ$ and no
systematic errors, the directions to the UHECR sources cannot be determined
with any precision because UHECR are likely to be charged particles and are
thus deflected in cosmic magnetic fields by potentially much larger
angles. These deflections are uncertain because of both the unknown particle
charge and uncertainties in the magnetic fields.

For their tiny flux, UHECR are only observed indirectly through extensive air
showers they produce in the atmosphere. This makes determining the nature (and
therefore the charge) of their primary particles prone to uncertainties of
hadronic interaction models. The existing measurements~\cite{Abbasi:2018nun, Abbasi:2018wlq, Aab:2016zth, Aab:2017cgk} have large
errors and may contain unknown systematic effects.

Cosmic magnetic fields are also not known sufficiently well. Experimentally, only loose bounds exist on the extragalactic fields: it is constrained by $10^{-15}$~G from below~\cite{Neronov:1900zz, Taylor:2011bn}
and by $10^{-9}$~G from above~\cite{Pshirkov:2015tua}. However, the arguments based on structure formation and measured 
fields in galaxy clusters indicate that the field in the voids should not
be much larger than $\sim 10^{-12}$~G, in which case the deflections of cosmic
rays in voids are negligible~\cite{Dolag:2004kp}, and sizable extragalactic deflections
may only arise either close to the source or close to our galaxy if it is itself a part of a filament~\cite{Courtois:2013yfa} with sizeable magnetic fields. 

A rough magnitude of the coherent Galactic magnetic field (GMF) is known to be several $\mu G$ from Faraday rotation measures of extragalactic sources and from other observations~\cite{2015ASSL..407..483H}. However, its
general structure is unknown because reconstruction of a 3d field from a 2d
projection is ambiguous. Several proposed phenomenological models
\cite{Han:2006ci, Sun:2007mx, Pshirkov:2011um, Jansson:2012pc} should be considered as examples of what the field might be, at
best. This makes it impossible to reconstruct the directions to the UHECR sources
with any certainty and trace where particular UHECRs have come from.

The arrival directions of observed UHECR events do not give any obvious indication of the nature of sources. 
The existing data appear quite isotropic, with no significant small scale
clustering found so far, and little large scale structure: there has been a
dipole of $6\%$ detected at intermediate energies of 8~EeV~\cite{Aab:2017tyv}, and a
concentrations of events --- the ``hot spots'' of the radius of $\sim
25^\circ$ --- at high energies above 57~EeV~\cite{Abbasi:2014lda}, with the significance
requiring further confirmation. Such remarkable isotropy, together with a
short propagation distance of all known charged particles at highest energies,
suggests deflections of at least a few tens of degrees for the bulk of UHECR
even at highest energies.

In the absence of a clear hypothesis of what UHECR are and where they come
from, one may opt for narrowing the search by excluding models. For this
approach to be useful three uncertainties --- sources, composition and
magnetic fields --- have to be somehow reduced to a manageable set by
additional assumptions. A most robust assumption can be made about the source
distribution in space: in all existing models they follow the matter
distribution. If one assumes in addition that the sources are sufficiently
numerous to be treated on statistical basis, the uncertainty related to sources is essentially
eliminated. We will refer to this source distribution as Large-scale Structure
(LSS) source model. This approach has already been used in previous
studies. For example, the lack of anisotropy in the data together with the
rough magnitude of GMF appears to be in tension with pure proton models~\cite{diMatteo:2017dtg}.

The unknown composition affects the distribution of arrival directions in two
ways: through the attenuation of UHECR, and through deflections in magnetic
fields. The situation here is under better control since for any type of
nuclei the attenuation can be calculated using available propagation
codes~\cite{Armengaud:2006fx, Aloisio:2012wj, Kalashev:2014xna},
so for any assumed spectrum and composition at the source the
spectrum and composition at the detector can be calculated. If not for the
uncertainty in magnetic fields, the sky distribution of the events would
have been calculable as well. Comparing observed and predicted sky
distributions would then allow one to constrain possible UHECR compositions or
even, if no good fit is found, to rule out the LSS source model.

In this paper we show that this logic largely survives the uncertainties of
the magnetic fields, and propose a method that allows one to constrain the 
charge composition of UHECR from a 
distribution of their arrival directions. To understand the idea imagine for
the moment that UHECR deflections were purely random. In this case they would
be characterized by a single parameter, the width of the Gaussian spread of a 
point source. To obtain the prediction
of a given model one could then calculate the distribution of 
arrival directions at zero magnetic field and smear it with the Gaussian 
function of a given width.
Comparing the result to observations one would determine/constrain the likely
values of the smearing angle and, given the rough magnitude of magnetic fields,  the composition
models.  

The real magnetic field is not random as there is a coherent field in the
Galaxy whose contribution to deflections is likely large, and which is
characterized my much more than one parameter even in simplest models.
However, one can still define a single observable which has a meaning of a 
typical deflection angle and which is robust to the presence of a regular
Galactic field in the sense of being insensitive to its details, but still
sensitive to the overall magnitude of deflections. We propose such an
observable below and investigate its discriminative power with respect to
different compositions of UHECR.

Recently, a number of studies have focused on disentangling an interplay between
cosmic rays anisotropy, mass composition and energy.
Following the observation of the dipole at intermediate UHECR energies by the Pierre Auger Observatory~\cite{Aab:2017tyv} and the subsequent indication of the growth of its amplitude with energy~\cite{Aab:2018mmi}, many theoretical works studied the implications of these signatures for UHECR sources and mass composition~\cite{diMatteo:2017dtg, Globus:2017fym, Wittkowski:2017nfd, Globus:2018svy, Mollerach:2019wne, Ding:2021emg}.
There are also several theoretical~\cite{Ahlers:2017wpb, Anjos:2018mgr, Erdmann:2018cvz, Kalashev:2019skq, Bister:2020rfv, Urban:2020szk, Wirtz:2021ifo} as well as observational~\cite{Aab:2014dha, Aab:2020mfn, Abbasi:2020fxl} studies that use more complex synthetic UHECR observables to identify the sources~\cite{Erdmann:2018cvz, Kalashev:2019skq, Bister:2020rfv, Urban:2020szk, Aab:2014dha, Aab:2020mfn, Abbasi:2020fxl} or infer mass composition~\cite{Ahlers:2017wpb, Anjos:2018mgr} and cosmic magnetic field~\cite{Wirtz:2021ifo}.
The approach rather close to ours was used in Refs.~\cite{Ahlers:2017wpb, Anjos:2018mgr}. For instance, in the study~\cite{Anjos:2018mgr} the likelihood method was developed to distinguish mass compositions of UHECR from specific sources.

The rest of this paper is organized as follows: in Section~\ref{sec:analysis}
we introduce the likelihood function and the test statistic (TS) that are the main analysis
tools of this study. In Section~\ref{sec:model} we describe the details of simulations of mock
UHECR event sets, including assumptions made about source distribution, UHECR composition, propagation and deflections in magnetic fields. 
In Section~\ref{sec:reconstruction} we check the accuracy and robustness of reconstruction of UHECR flux parameters with this TS.
Making use of the simulated TS distributions,  in Section~\ref{sec:results} we formulate and test the method to constrain the UHECR mass composition and to compare different composition models. We also discuss the impact of magnetic field uncertainties and energy threshold variation on the results.
We present our conclusions in Section~\ref{sec:discussion}.

\section{The choice of test statistics and observable}
\label{sec:analysis}

The key ingredient of our proposal is the choice of the test statistics and 
the corresponding observable. To distinguish between different compositions, we
want it to depend on the overall magnitude of deflections but be insensitive to
their particular directions. One may expect that such observable will not
depend strongly on the details of the coherent magnetic field, but mainly on its overall magnitude --- the latter is the parameter best known from observations. Note that the existing GMF models agree on the 
overall magnitude of the Galactic field within $\sim 50$\%, which has smaller effect on deflections than the
uncertainty in the particle charge which ranges from 1 for protons to 26 for
iron. One may thus expect to constrain the composition despite the relatively poor
knowledge of the magnetic field.

Our choice of observable is inspired by the case of purely random deflections which are characterized by a single parameter, the width of the Gaussian spread of a point source. 
By analogy, we choose to characterise the given set of arrival deflections by its typical deflection angle  with respect to the LSS source model. 
Given the set of events, this quantity is calculated as follows. 

For a given smearing parameter $\theta$ we construct the sky map of the expected flux
making use of the source distribution in space and the exposure of the
experiment (the procedure is described in detail in Sec.~\ref{sec:model}). We
characterize this flux by a flux map  $\Phi(\theta, {\bf n})$ --- a continuous
function of the direction, which is normalized to a unit integral over
the sphere so that it can be interpreted as a probability density to observe
an event from the direction ${\bf n}$.

Given the flux map $\Phi(\theta, {\bf n})$ it is straightforward to generate
the set of events that follow the corresponding distribution by throwing
random events and accepting them with the probability $\Phi(\theta, {\bf n})$
according to their direction ${\bf n}$. Inversely, given the set of events with directions ${\bf n}_i$ generated with some value of $\theta$, one can determine the value of $\theta$ by computing the $\theta$-dependent likelihood function
\begin{equation}
\mathcal{L}(\theta) = \sum_i \ln  { \Phi(\theta, {\bf n}_i) \over 
\Phi_{\rm iso} ({\bf n}_i)}  
\label{eq:likelihood}
\end{equation}
and finding its maximum with respect to $\theta$. Here for convenience we
have chosen the normalization factor $\Phi_{\rm iso}({\bf n}_i) = \Phi(
\infty,{\bf n}_i)$ that corresponds to the isotropic distribution of sources ---
a uniform flux modulated by the exposure function.

So far we have assumed that all CR events are deflected in the same way, while
in a realistic situation the events have different energies. Accounting for the
energy dependence does not introduce additional parameters as the deflection
angles are inversely proportional to event energies and can be expressed in
terms of a single parameter, e.g. the deflection at a reference energy
$E_0=100$~EeV. We bin the energies in log-uniform intervals with lower
boundaries $E_k$ (ten bins per energy decade with the highest bin an open interval $E > 180$~EeV) and neglect the energy dependence within each bin.
 We then
define a flux map $\Phi_k(\theta,{\bf n})$ in each energy bin. Note that the
attenuation of cosmic rays is energy-dependent, so the flux maps
$\Phi_k(\theta,{\bf n})$ at different $k$ differ not only by the deflection
angle, and in general
$\Phi_k(\theta, {\bf n}) \neq \Phi_i(\theta E_i/E_k, {\bf n})$.

Generalizing Eq.~(\ref{eq:likelihood}) to the case of several energy bins, we
finally define our test statistics $TS(\theta)$ as follows:
\begin{equation}
TS(\theta) = -2 \sum_k \left( 
\sum_i \ln  { \Phi_k(\theta, {\bf n}_i) \over 
\Phi_{\rm iso} ({\bf n}_i)}
\right), 
\label{eq:TS}
\end{equation}
where the internal sum runs over the events in the energy bin $k$ and we have
included a standard normalization factor $-2$. In the limit of a large number of
events, this test statistics is distributed around its minimum according to
$\chi^2$-distribution with one degree of freedom.

By definition, the test statistics $TS(\theta)$ is calculated using the LSS
source model. As already mentioned, it also implicitly depends on the
attenuation of cosmic rays. We adopt the proton attenuation function when
calculating the flux maps $\Phi_k(\theta,{\bf n})$. In
principle this choice is arbitrary. Note, however, that protons attenuate less
than other particles, so the maps $\Phi_k(\theta,{\bf n})$ calculated with
proton attenuation are closer to being isotropic and thus this choice is
conservative. 

On the contrary, Eq.~(\ref{eq:TS}) makes no reference to any
magnetic field model. Moreover, its sensitivity to the coherence of the field
is reduced: for instance, for a single point source the value of $TS(\theta)$
is the same for events spread on the circle around the source as would be the
case for random deflections, or events concentrated in one point on the
circle as in the case of a coherent field.

The observable we propose  to characterize a given set of events is the
value of $\theta$ for which the $TS(\theta)$ is minimum. 
In what follows we call this value $\theta_{\rm rec}$ for ``reconstructed''. It has the
interpretation of the typical deflection angle with respect to the LSS source 
model at the reference energy $E_0=100$~EeV. Its uncertainty is determined by 
the width of the minimum.

\section{Calculation of test statistics and modeling of UHECR flux}
\label{sec:model}

We now have to calculate the test statistics defined in Eq.~(\ref{eq:TS}). We also want to test 
its behavior for different compositions and magnetic field models, and therefore we need to generate 
Monte-Carlo event sets that follow these models. Both problems are solved by computing corresponding
flux density maps. The goals, however, are different: in the case of test statistics we want to keep it as simple and model-independent as possible, while in the case of test UHECR sets we want to be as close to reality as we can achieve with available computational resources. 
The general steps are as follows. First, we model the source
distribution in space and compute the flux, as a function of direction,  as it
would be observed at the borders of the Galaxy at a given energy $E_k$. The UHECR attenuation enters
at this stage, but not the deflections. At the second step we add the
deflections. To get the maps $\Phi_k(\theta,{\bf n})$ entering the test statistics Eq.~(\ref{eq:TS}) we apply a simple Gaussian smearing of width $\theta$.  In the case of flux maps used to generate model event sets,
 we apply a latitude-dependent smearing as described in the next Section and additionally process the flux through the regular Galactic magnetic field. To avoid confusion, we denote these maps as $F_k$. 
Finally, we use the model maps $F_k$ to generate the test event sets which we need to study the behavior of the test statistics Eq.~(\ref{eq:TS}). We detail below these steps.

\subsection{UHECR sources}
\label{sec:sources}

As stated in the Introduction, we assume that the sources follow
the large-scale matter distribution in the Universe, and that they are
sufficiently numerous to be treated on statistical basis. Specifically, we
assume that much more than one source is present in galaxy clusters and larger structures.

Quantitatively, this means that the source density must be much larger than $\sim
10^{-5}$~Mpc$^{-3}$.

While this is true in many models,
there are source candidates that are more rare (for
instance, powerful radio-galaxies) and for which this assumption does not
hold. Note that the latter case physically corresponds to the situation when
the UHECR propagation horizon, which is of order of a few tens of Mpc at
highest energies, contains one or a few sources only.

We estimated the applicability of our method in this situation and found
that for rare sources a sizable anisotropy would likely be observed
at present UHECR statistics even assuming a heavy composition, see Section~\ref{sec:results} for details. As such an anisotropy is in tension with the existing UHECR data, the assumption of numerous 
sources appears reasonable.

This assumption, together with the known distribution of galaxies up to
a distance of $200-300$~Mpc, provides one with all the necessary information
about the space distribution of UHECR sources.

Consequently, the UHECR flux at Earth
predicted in this model depends only on the CR composition (affecting both 
cosmic ray propagation and deflections) and on magnetic fields.

In order to reconstruct the source distribution in space (and therefore, on
the sky) under the above assumptions, we assign each galaxy an equal {\em intrinsic}
luminosity in UHECR.  In practice, we use the 2MRS galaxy catalog \cite{Huchra:2011ii}
which contains galaxy distances. We cut out dim galaxies with ${\rm mag} >
12.5$ so as to obtain a flux-limited sample with a high degree of
completeness, and eliminate galaxies beyond $250$~Mpc. We assign progressively
larger flux to more distant galaxies to compensate for the observational
selection inherent in a flux-limited sample (see Ref.~\cite{Koers:2009pd}  for the exact
procedure). In a similar way, we give more weight to the galaxies within $\pm
5^\circ$ from the Galactic plane to compensate for the catalog
incompleteness in this region. Finally, we assume that sources beyond
$250$~Mpc are distributed uniformly with the same mean density as those within
this distance. We obtain in the end the space distribution of sources that is
completely fixed.

\subsection{UHECR propagation and deflection in magnetic fields}
\label{sec:propagation}

\begin{figure*}
\begin{center}
 \includegraphics[width=0.49\columnwidth]{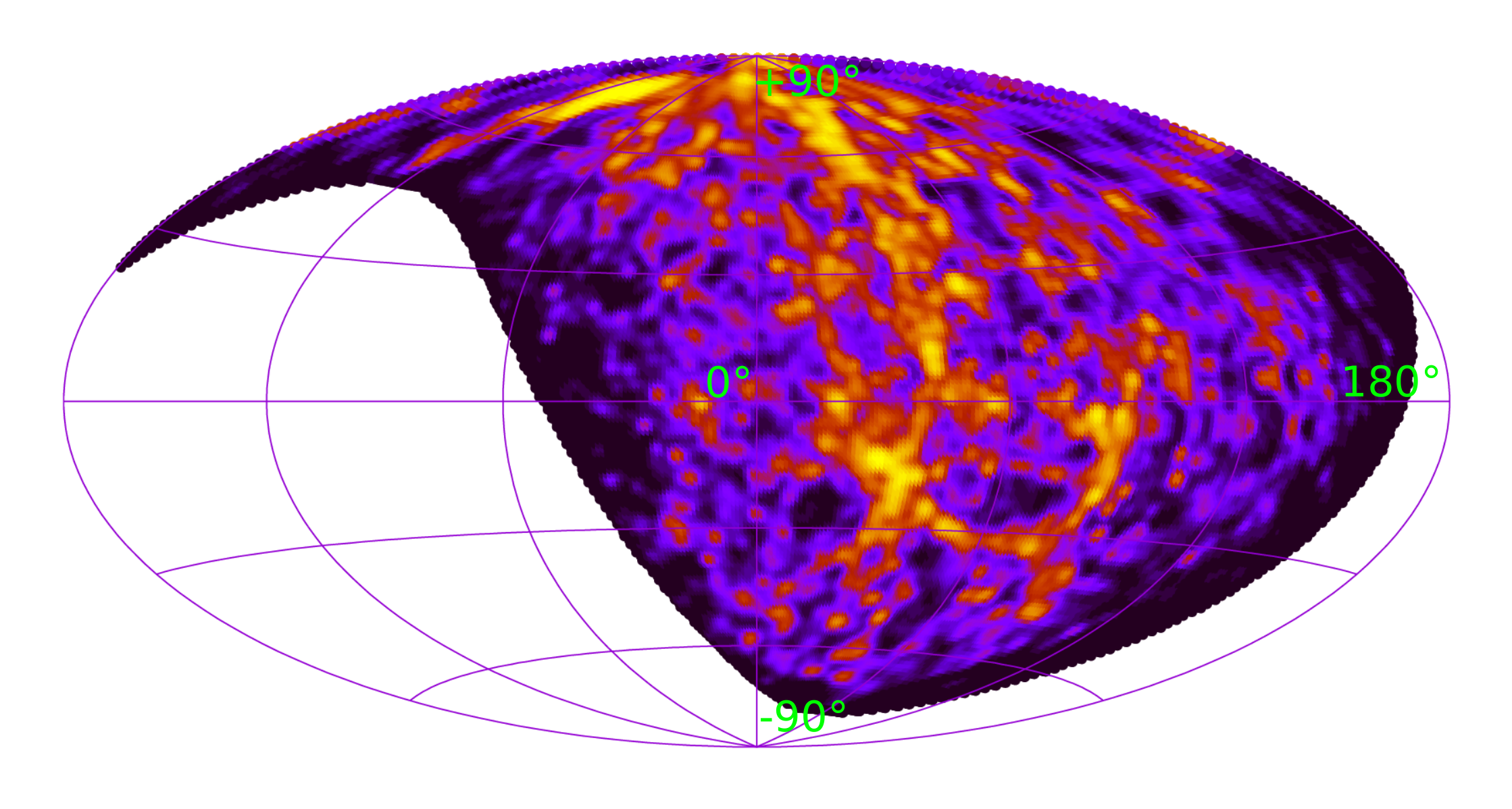}
 \includegraphics[width=0.49\columnwidth]{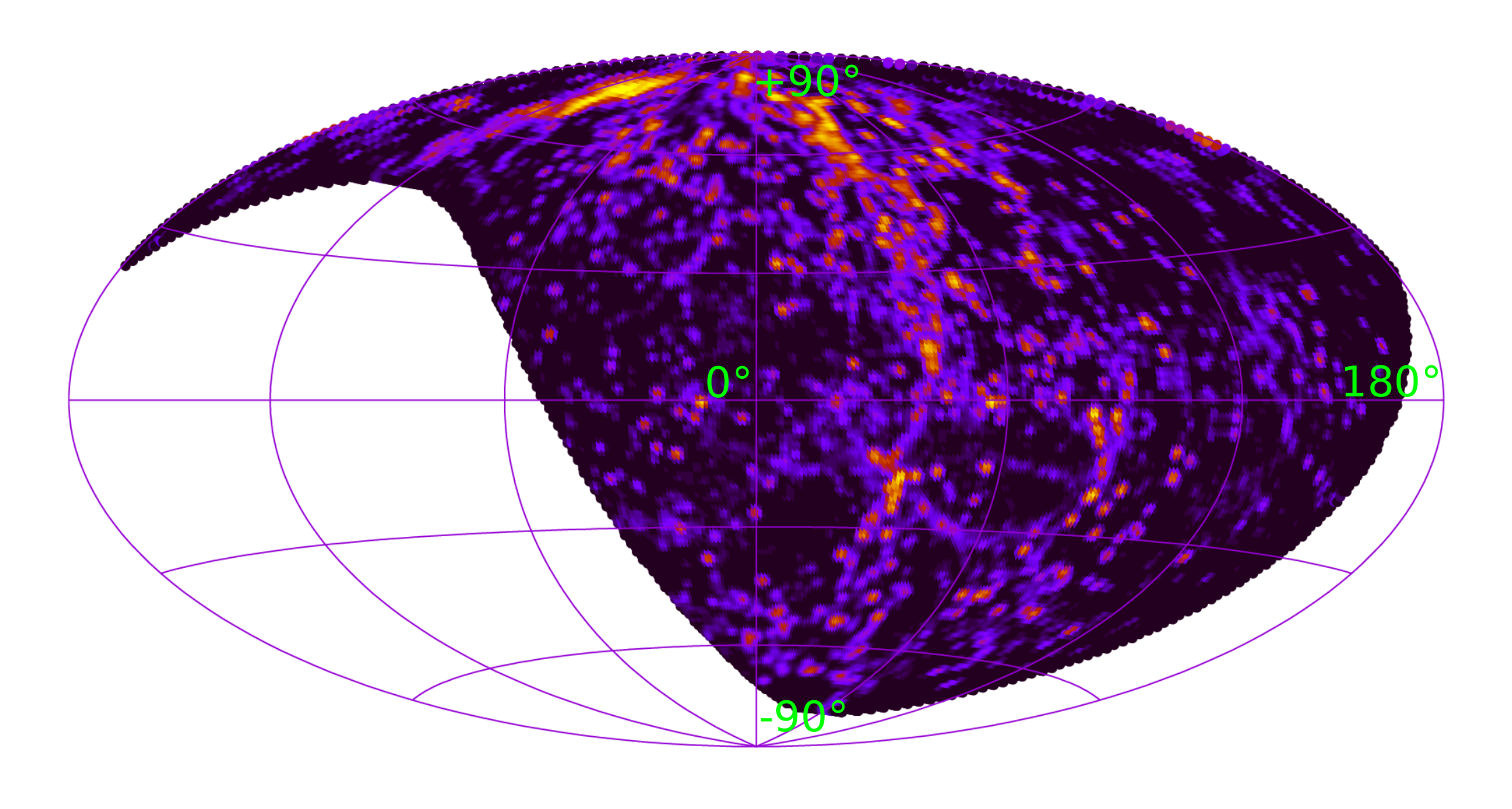}
 \includegraphics[width=0.49\columnwidth]{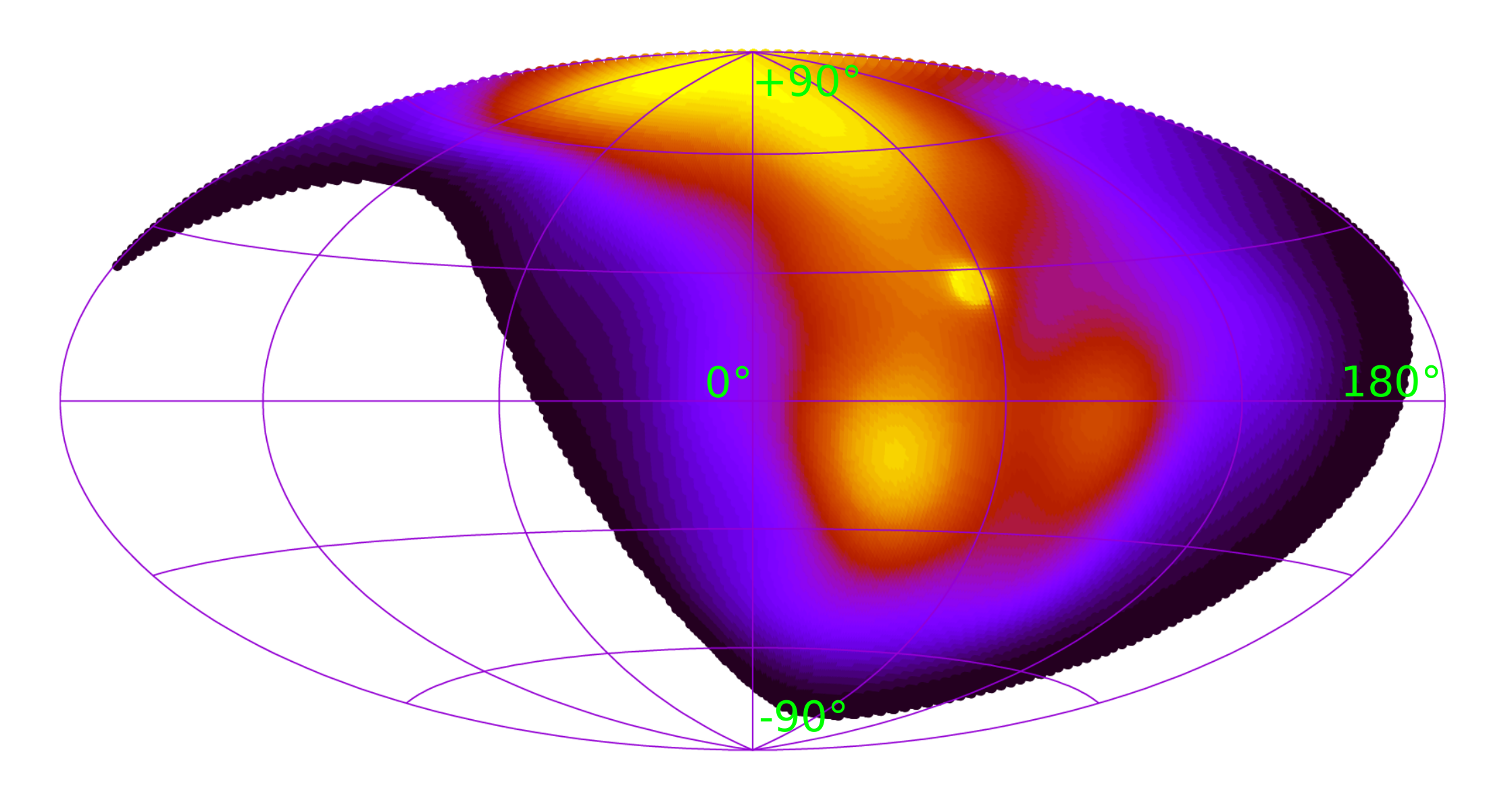}
 \includegraphics[width=0.49\columnwidth]{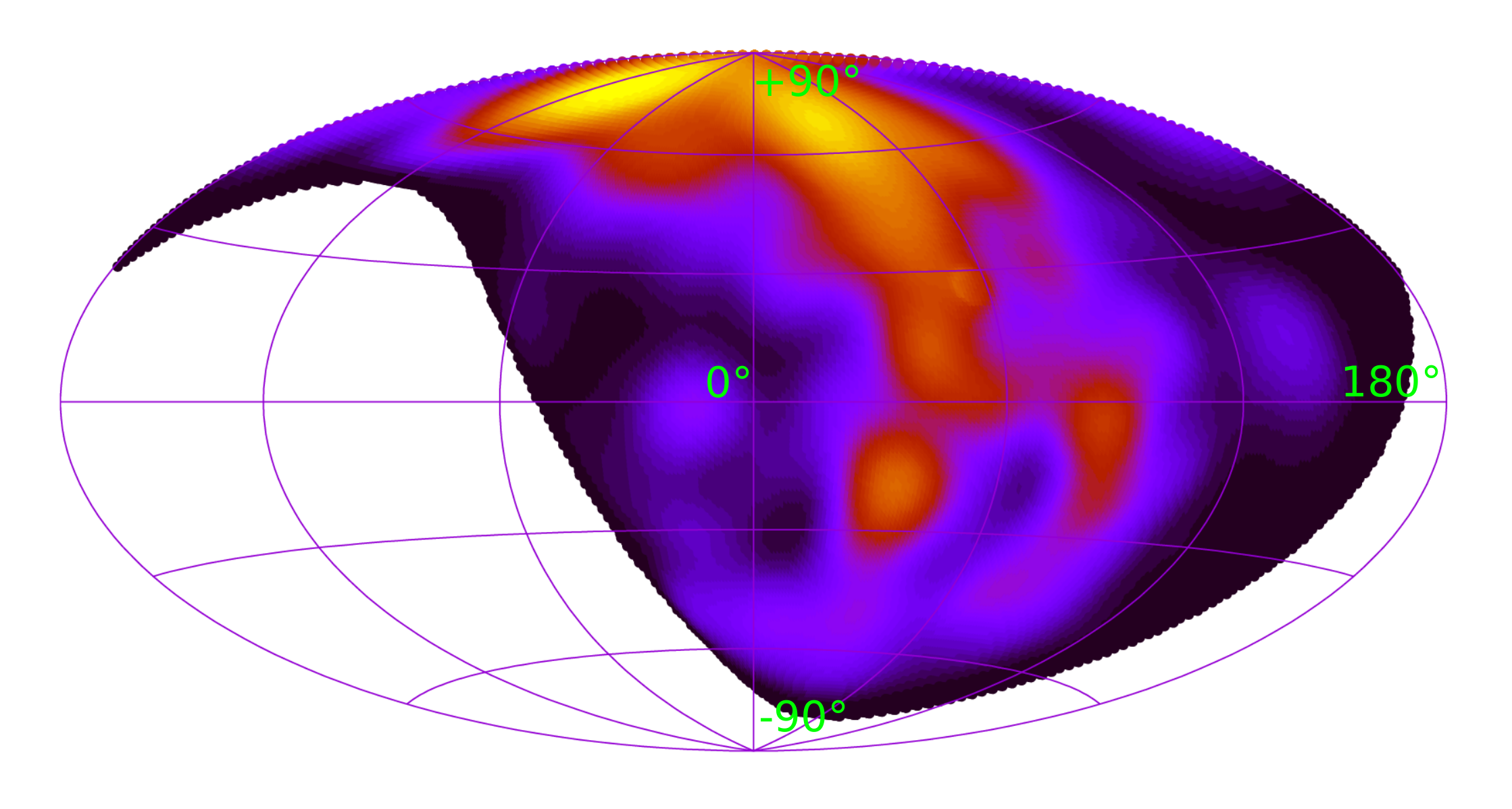}
\caption{
\label{fig:maps_TS}
Examples of UHECR flux model maps $\Phi_k$ used for test statistic calculation. Proton attenuation length and uniform smearing of the sources is assumed: $\theta(100\; {\rm EeV}) = 1^\circ$ (top) and $\theta(100\; {\rm EeV}) = 10^\circ$ (bottom). Energies are $E_k = 57$~EeV (left) and $E_k = 100$~EeV (right).
Maps are shown in galactic coordinates for TA SD field of view.
}
\end{center}
\end{figure*}

Contributions of individual sources to the observed flux are affected, apart
from the trivial $1/r^2$ falloff, by the attenuation and deflections in
magnetic fields. In practice these two effects can be separated. Most of the
attenuation happens outside of the Galaxy where deflections are random
and probably negligible all together, while in the Galaxy the deflections are
important but the attenuation is negligible. We therefore calculate first, by
making use of SimProp~v2r4 code~\cite{Aloisio:2017iyh}, the attenuated but non-deflected flux in a given energy bin as it arrives to the Galaxy borders. 
The flux in the energy bin $k$, calculated with the proton attenuation and
smeared with the Gaussian function of the width $\theta$, gives the flux map
$\Phi_k(\theta,{\bf n})$ which enters the definition of our test statistics,
Eq.~(\ref{eq:TS}). We show several examples of these maps in Fig.~\ref{fig:maps_TS}.

\begin{figure*}
\begin{center}
 \includegraphics[width=0.49\columnwidth]{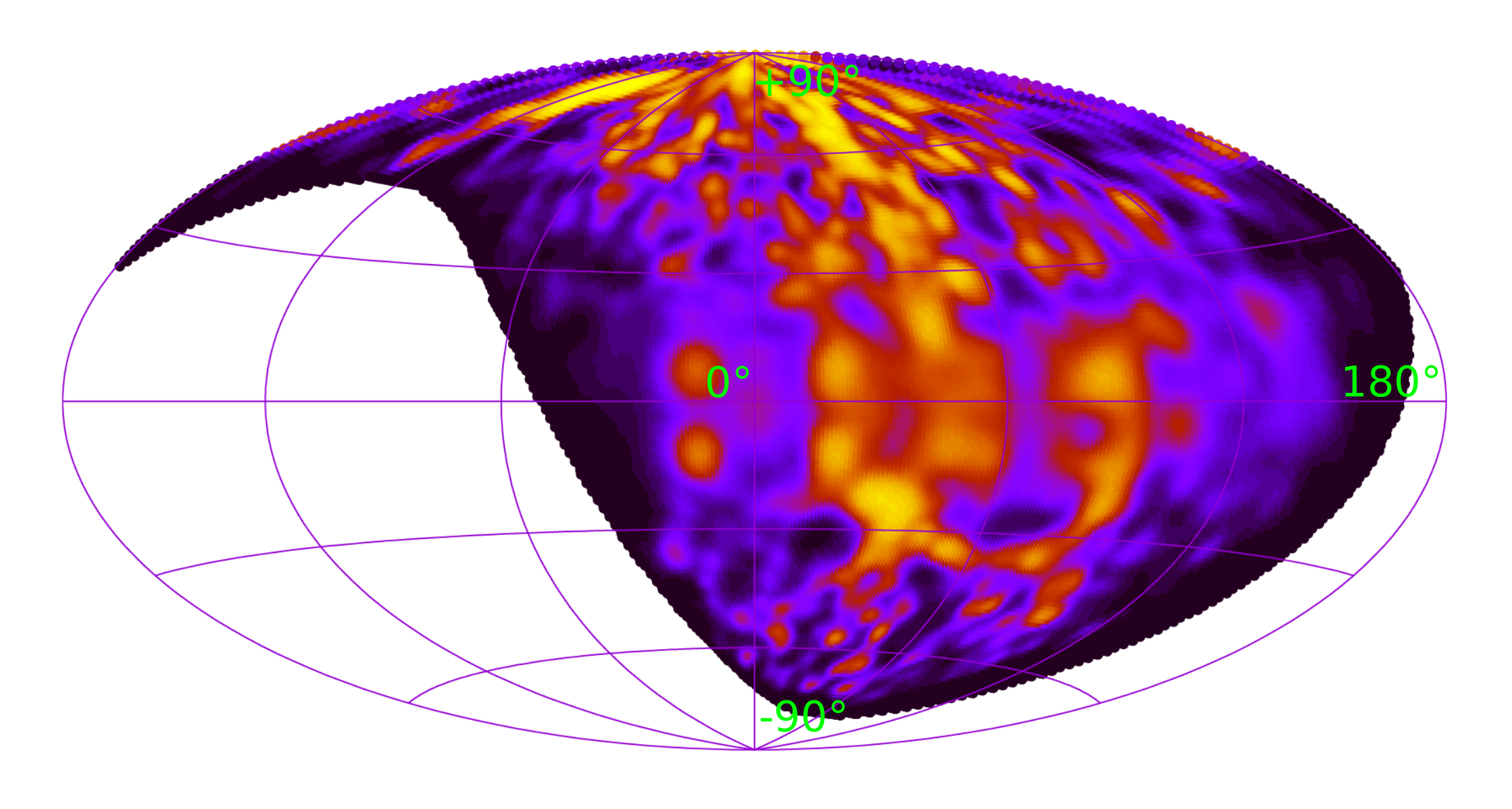}
 \includegraphics[width=0.49\columnwidth]{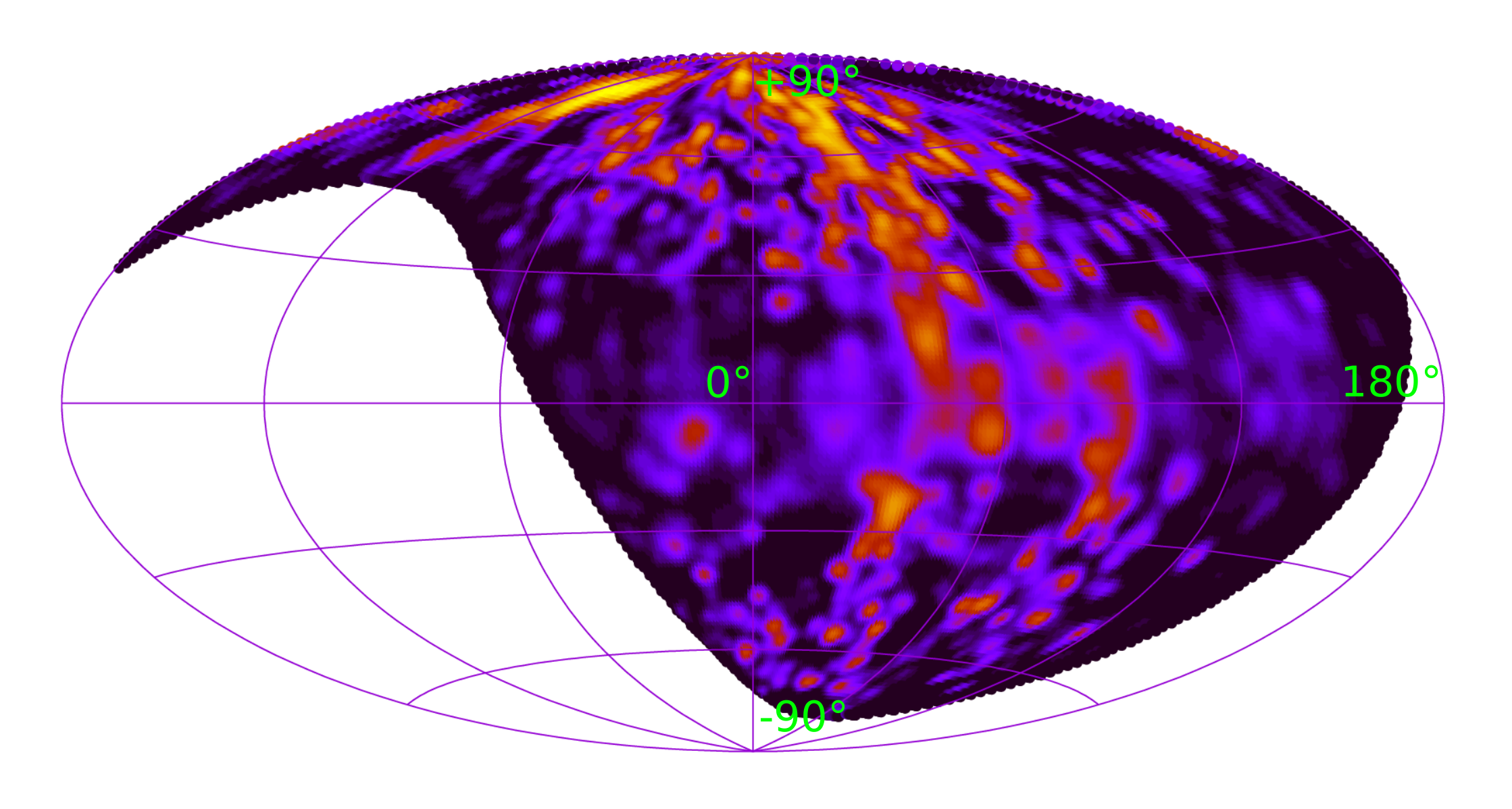}
 \includegraphics[width=0.49\columnwidth]{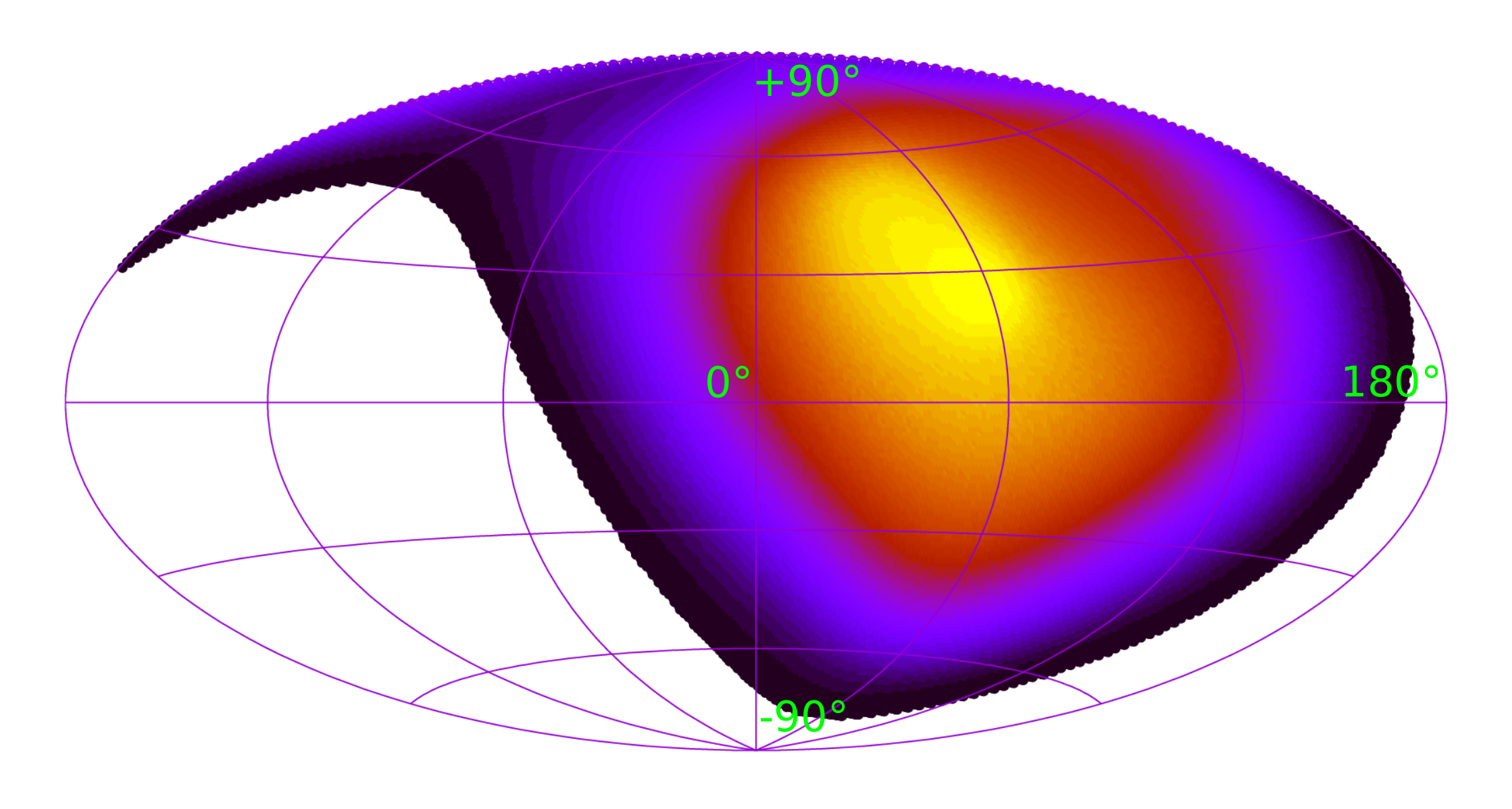}
 \includegraphics[width=0.49\columnwidth]{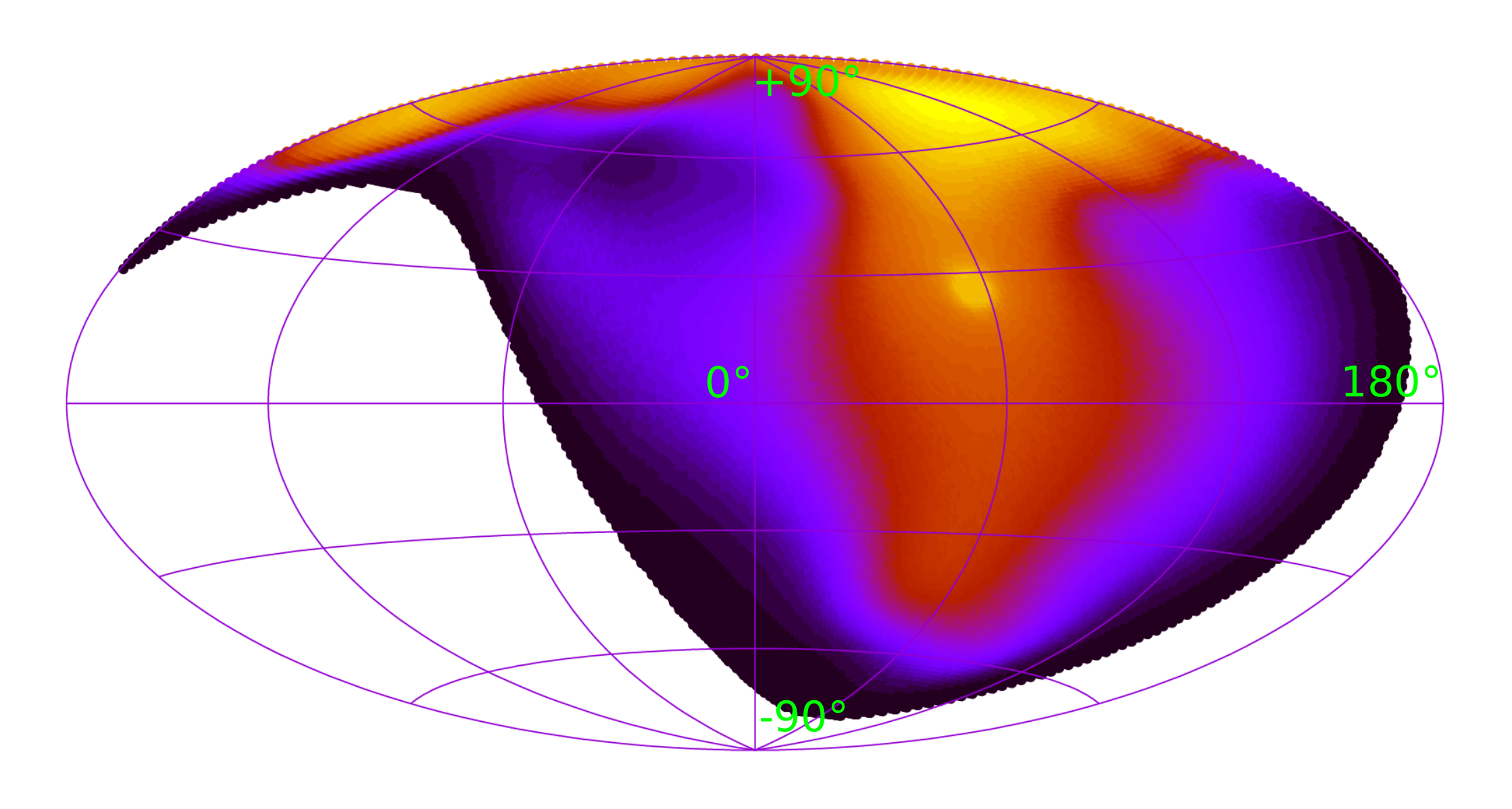}
\caption{
\label{fig:maps_model}
Examples of UHECR flux model maps, $F_k$ used for mock UHECR sets simulation. 
Maps for protons with b-dependent smearing of the sources and {\it regular GMF effect}: $E = 57$~EeV (top left) and $E = 100$~EeV (top right).
Maps for iron with b-dependent smearing of the sources and {\it regular GMF effect}: $E = 57$~EeV (bottom left) and $E = 100$~EeV (bottom right).
Maps are shown in galactic coordinates for TA SD field of view.
}
\end{center}
\end{figure*}

For the generation of test event sets we need flux maps
computed with the attenuation corresponding to different compositions. 
While straightforward in principle, the calculation of attenuation in the
general case is fairly complicated as different species have different
attenuation lengths and produce different sets of secondaries. Leaving the
general treatment for future study, we simplify the problem by restricting
ourselves to a proton-iron mix as the injected composition. This choice has
two advantages. First, the proton-iron mix can be treated in the
attenuation-only approximation, while secondaries can be neglected. Indeed,
protons produce no secondaries (except very small number of gamma-rays and
neutrinos that do not affect the general flux
picture~\cite{Gelmini:2007jy, Gelmini:2011kg, AlvesBatista:2019rhs}).
For iron nuclei the attenuation length
is larger than for all its secondaries~\cite{Puget:1976nz, Lee:1996fp, diMatteo:2020dlo}.
Furthermore, secondary protons from iron propagation
have a factor $\sim 50$ lower energies than the primary nuclei and
drop out from the high-energy range $E\gsim 10$~EeV that we consider in what
follows. Second, the iron nuclei are deflected in magnetic fields 
stronger than all their secondaries, which makes the event distribution more
uniform, and therefore conclusions based on anisotropies are conservative.  
The attenuation curves for protons and iron are obtained by fitting the results of
SimProp~v2r4~\cite{Aloisio:2017iyh} simulations in the same way as in Ref.~\cite{diMatteo:2017dtg}.
Note that we do not vary the injection spectrum of UHECR while calculating the flux maps, rather we fix it to the power-law with the spectral index $\alpha = -2.5$ and no cut-off. At the same time we assume the simulated mock event sets to follow the observed spectrum, see Sec.~\ref{sec:mock_sets}. While this is a reasonable approximation, it is not entirely self-consistent: in the exact calculation the injection spectrum should be fitted to the observed one within a given composition model. We think, however, that for the estimate of the sensitivity of the method our approximation is sufficient. 

The next step, which is only needed for the model maps $F_k$, is correcting the 
flux that arrives at the borders of the
Galaxy by the deflections in the Galactic magnetic fields. The latter has
regular and random components.  For the regular field we take one of the two
models~\cite{Pshirkov:2011um, Jansson:2012pc}. Both models give the regular
magnetic field everywhere in the Galaxy as a function of a large number of
parameters whose values are determined by fitting to the observational
data. In each case we adopt the best fit values of these parameters as given
in Refs.~\cite{Pshirkov:2011um, Jansson:2012pc}. Given the magnetic field, we
convert the flux density outside of the Galaxy into the observed flux density
as follows. To determine the observed flux density in a given direction ${\bf
  n}$ for a given species of charge $q$ and given energy bin we back-track a
particle of charge $-q$ and corresponding energy launched in the direction
${\bf n}$ through the galactic magnetic field until it leaves the Galaxy in
some other direction ${\bf n'}$. The observed flux density in the direction
${\bf n}$ is given by the external flux density in the direction ${\bf
  n'}$. The total flux map in a given energy bin is the weighted sum of maps 
for individual species.

The deflections in random magnetic fields, both Galactic and extragalactic, 
have an effect of smearing of the
flux density. The smearing is proportional to the combination $Bq/E$ and is
different for different UHECR species and energies $E$. If the random magnetic
field was direction-independent the smearing would have been uniform over the sky. This is what we have adopted for the definition of the test
statistics.  Because of the presence of the Galactic random field which
has a space-dependent magnitude, in reality the random deflections 
depend on the
direction. We take this effect into account when generating the 
test event sets. The dependence of mean deflections $\sqrt{ \langle
  \theta^2\rangle}$ (equivalently, the smearing angle) on the Galactic latitude has been
estimated from the dispersion of Faraday rotation measures of
extragalactic sources in Ref.~\cite{Pshirkov:2013wka} where the following
empiric relation has been obtained for protons of $E=40$~EeV:
\begin{equation}
\sqrt{ \langle \theta^2\rangle} \leq {1^\circ\over \sin^2b +0.15}
\label{eq:smearing}
\end{equation}
We conservatively adopt this relation treating it as the equality (i.e,
assuming maximum deflections). 
 
For other species and energies we rescale the deflections
according to magnetic rigidity. 
Note that this relation is just a phenomenological parameterization of the 
dependence of random deflections on the Galactic latitude; no clear dependence on the Galactic longitude has been detected in Ref.~\cite{Pshirkov:2013wka}.

When the smearing is non-uniform and not small, there is a subtlety in how to
implement it in a way that is accurate enough and not too complicated. A
technically simplest option --- to apply the full smearing by a latitude-dependent angle $\theta$
at once --- is inaccurate at large gradients of $\theta$ as it would convert a point source
into a circular distribution while in reality a deformed one is expected. We apply instead  a series of $N$ (in practice $N \sim 10$) {\it smaller
identical smearings} where 
the smearing angle at one step $\sim
\theta/\sqrt{N}$ is normalized in such a way that the result is identical to the one-step smearing by $\theta$ in the direction-independent case.

The deflection of UHECR in extragalactic magnetic fields can be estimated as~\cite{Bhattacharjee:1998qc}:
\be
\label{eq:EGMF}
\Delta \theta \simeq 0.8^\circ \cdot \frac{100~{\rm EeV}}{E} \cdot \frac{B}{1~{\rm nG}} \cdot \sqrt{\frac{L}{10~{\rm Mpc}}} \cdot \sqrt{\frac{\lambda}{1~{\rm Mpc}}} \cdot Z
\ee
where $E$ and $Z$ is UHECR energy and charge respectively, $B$ and $\lambda$ is EGMF strength and correlation length, respectively, and $L$ is the distance traversed by the UHECR. As already mentioned in Sect.\ref{sect:Introduction},  there is little observational knowledge about the strength and structure of the extragalactic magnetic fields. Upper~\cite{Pshirkov:2015tua} and lower~\cite{Neronov:1900zz, Taylor:2011bn} limits exist for the field strength in voids,  and only upper limits for the fields in filaments~\cite{Brown:2017dwx, Locatelli:2021byc}~\footnote{ See however a recent study about the detection of magnetic field of a distant filament~\cite{Vernstrom:2021hru}.  }.
The field correlation length is even more uncertain, with no direct observational bounds~\cite{Durrer:2013pga}. 

Given the observational uncertainties, in subsequent discussion we resort to numerical simulations~\cite{Hackstein:2016pwa, Hackstein:2017pex, Garcia:2020kxm}. The latter show that the strength and structure of the field in voids and super-galactic structures may vary significantly depending on the model of the EGMF origin and on the simulation setup, reaching the values which may potentially impact our study in two cases: large fields in voids and strongly magnetized local filament (by local we mean a filament {\em containing} the local group). 

Concerning the fields in voids, they may impact our results if either the filed is at the highest value allowed by observations  $B \gtrsim 1$~nG and $\lambda \gtrsim 200$~kpc, or for fields $B \gtrsim 0.1$~nG and $\lambda \gtrsim 20$~Mpc. Both cases are rather extreme. 
Alternatively, our Galaxy itself may be situated in a magnetized filament. The magnetized regions in general follow the LSS and hence our assumed sources distribution. Therefore any deflection that occurs on a distant magnetized structure would only mimic the source located in that region and cannot spoil our general flux picture.
The recent constrained simulation of EGMF in the local Universe~\cite{Hackstein:2017pex} indicates the presence of a $\sim 5$~Mpc large local filament around the Milky Way magnetized to $\sim 0.3 - 3$~nG over most of its volume in most pessimistic case. The impact of this structure on UHECR deflections would supersede that of GMF only if its field is highly coherent with $\lambda \gtrsim 0.2 - 5$~Mpc. 

From these arguments, it is reasonable to neglect the impact of EGMF on UHECR deflections in our flux model. However, this possibility cannot be excluded. If deflections due to EGMF are not negligible, our method is still applicable though it may be less sensitive. Regardless of their origin, the extragalactic deflections are characterized by a single parameter, a typical deflection angle. In what follows we present our results assuming these deflections are unimportant, and estimate the value of this parameter at which this assumption breaks down.

Finally, we apply the additional uniform smearing by $1^\circ$ 
to account for experimental angular resolution. 
It only slightly affects proton maps at high energies in the Galactic pole regions
where the deflections due to random and regular magnetic fields are smaller
than $1^\circ$.  Note that if random deflections in the extragalactic fields are non-negligible they can also be added at this step.  Several examples of resulting model flux maps
$F_k$ are shown in Fig.~\ref{fig:maps_model}. 

When we study the sensitivity of our method below in Secs.~\ref{sec:reconstruction},\ref{sec:results} we generate
model flux maps using the regular GMF of Ref.~\cite{Pshirkov:2011um},
the random GMF given by Eq.~(\ref{eq:smearing})  and neglecting EGMF. 
In order to test the dependence of the sensitivity on the magnetic field parameters we also include in Sec.\ref{sec:uncertainties} the field of Ref.~\cite{Jansson:2012pc}.
For both models, we vary independently the overall magnitudes of regular and random fields

and introduce the uniform smearing in EGMF.

\subsection{Generation of mock UHECR event sets}
\label{sec:mock_sets}
Given model flux maps $F_k$, it is straightforward to generate a mock set
of UHECR events as it would be detected by an EAS experiment at Earth. We modulate the
all-sky flux map calculated as explained in Sec.~\ref{sec:propagation}
by the exposure function of the TA experiment, for which we take the geometrical
exposure. Once multiplied by the exposure and normalized to a unit integral,
the flux map is interpreted as probability distribution for the arrival
directions, so the latter can be generated directly by throwing random events
and accepting them with the probability given by the corresponding flux map 
at the position of the event.

The energies of events are generated randomly according to the actual TA SD
spectrum~\cite{Ivanov:2015pqx}.  Throughout the main part of this study we set
the lower energy threshold of the mock events to be $E_{\rm min} =
57$~EeV, thus aiming to study the properties of UHECR in GZK-cutoff
region. Note that this threshold energy is also adapted in the anisotropy studies by TA
SD. In Section~\ref{sec:low_energy} we consider lower energy thresholds
down to $E_{\rm min} = 10$~EeV and discuss the effect this has on the
results.

\section{Reconstruction of flux parameters with the test statistics}
\label{sec:reconstruction}
Before we can assess the sensitivity of our method in realistic cases it is instructive to check how it works for mock event sets generated with the same maps $\Phi_k$ as used in the definition of the test statistics (\ref{eq:TS}), i.e. with uniform smearing only and without
regular magnetic field effects. If we generate a mock event set with a given smearing parameter  $\theta_{\rm th}$, we should recover the value of $\theta_{\rm th}$ by calculating the test 
statistics $TS(\theta)$ and finding its minimum. In the limit of large number of events $N_{\rm ev}$ the test statistic $TS(\theta)$ should follow the $\chi^2$ distribution with one degree of freedom.

\begin{figure*}
\begin{center}
 \includegraphics[width=0.49\columnwidth]{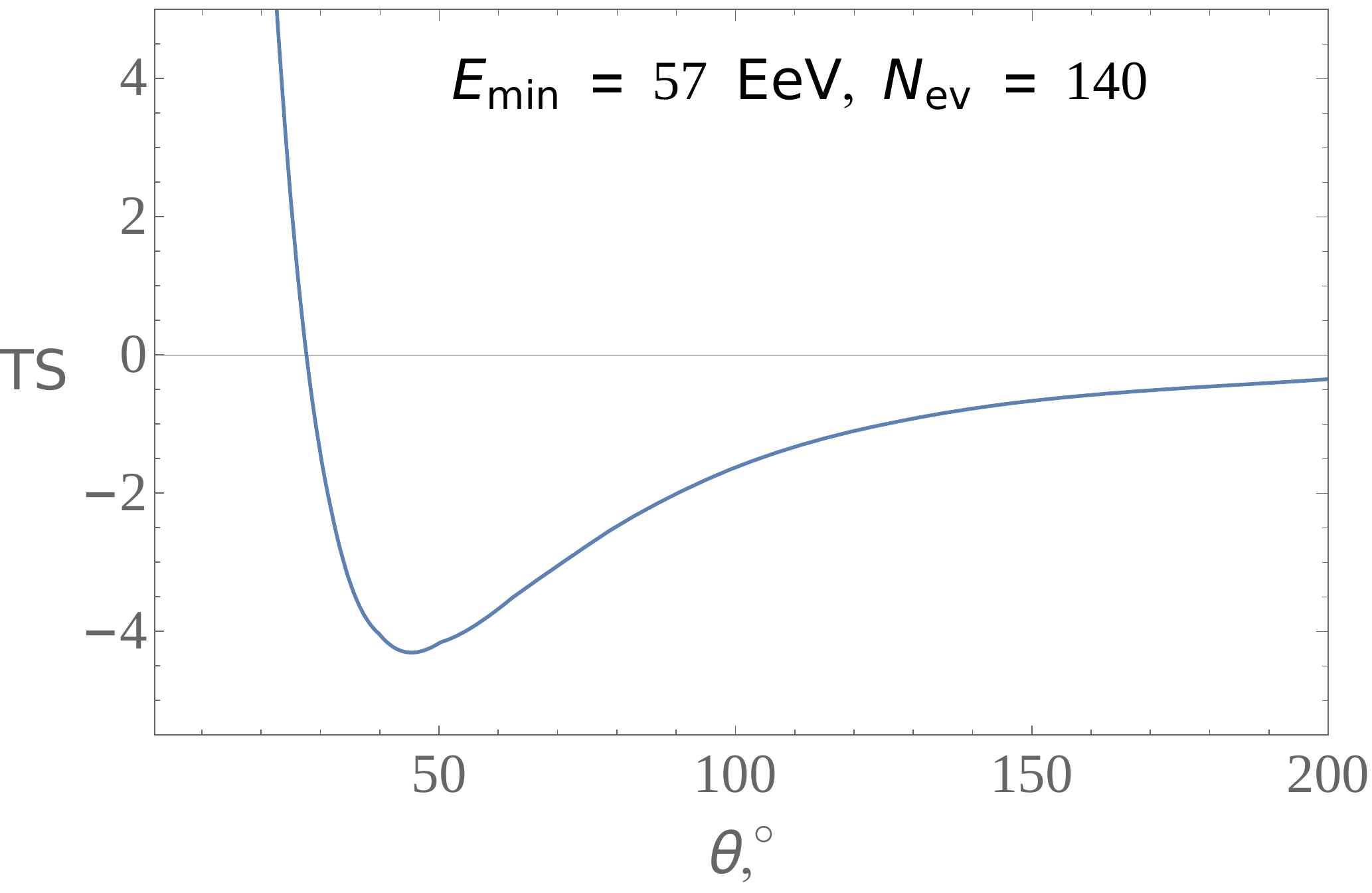}
 \includegraphics[width=0.49\columnwidth]{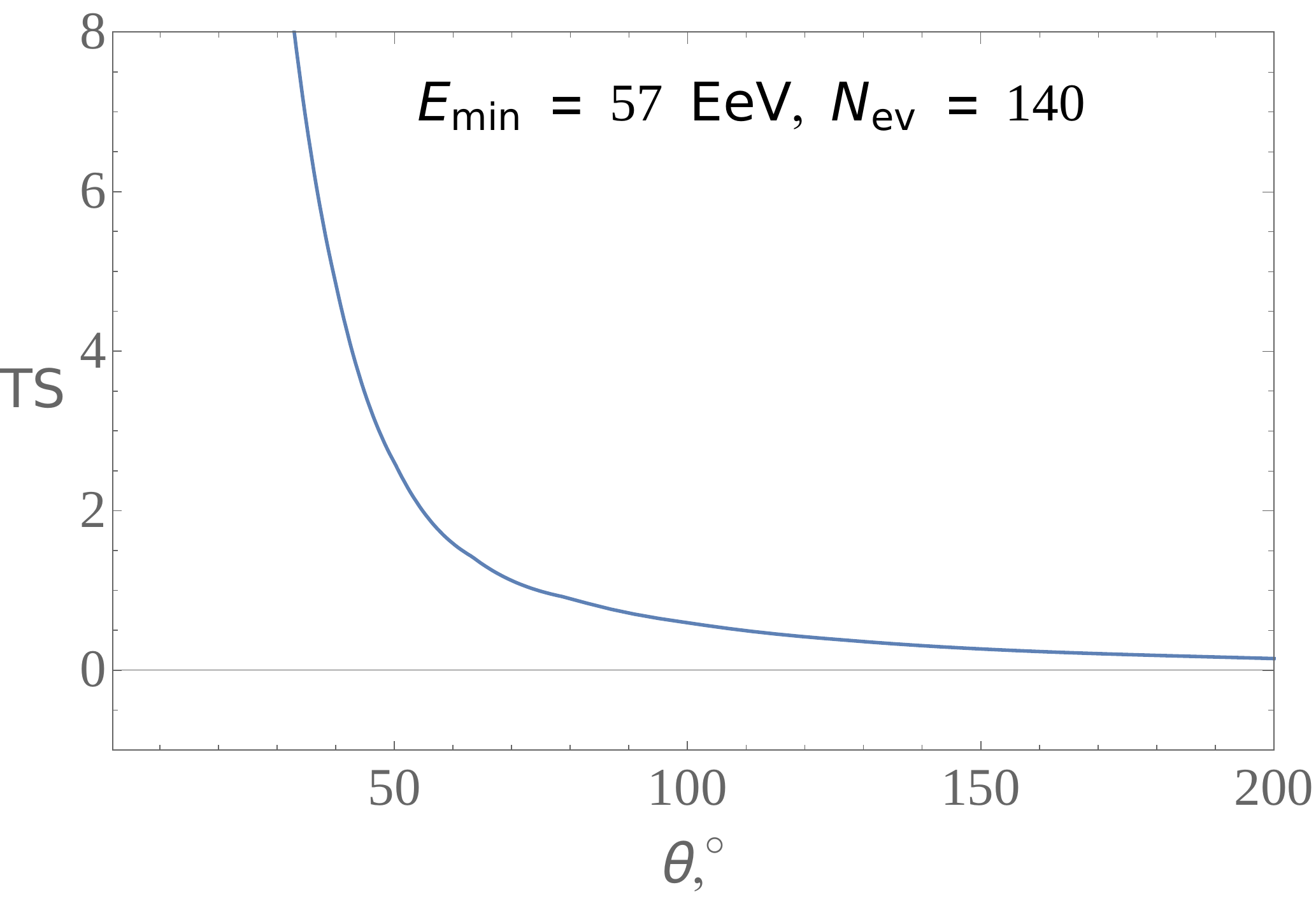}
\caption{
\label{fig:LLH_samples}
Examples of TS distribution for two different event sets with the same parameters:
$E_{\rm min} = 57$~EeV and $N_{\rm ev} = 140$. {\it Left:} $TS$ has a global minimum:
$\theta_{\rm rec} = 45.4^\circ$. {\it Right:} $TS$ has no global minimum: $\theta_{\rm rec} = 200^\circ$.
}
\end{center}
\end{figure*}

To check how well this picture is reproduced for a finite number of events in a set
we generate a large number of sets with $N_{\rm ev}$ events, and calculate $TS(\theta)$ for each set. We record the position of the minimum $\theta_{\rm rec}$ and the width  $\Delta \theta_{\rm rec}$ at 
$TS(\theta) = TS_{\rm min} + 1$ as corresponds to $1\sigma$ interval for the $\chi^2$-distribution. 
If the minimum is not found in the range $0 \leq \theta < 200^\circ$ 
we conclude that the given event set cannot be distinguished from an isotropic one by our $TS$
and assign it a value $\theta_{\rm rec} = 200^\circ$. 
Two examples of $TS(\theta)$ are shown in Fig.~\ref{fig:LLH_samples}. 

\begin{figure*}
\begin{center}
 \includegraphics[width=1.00\columnwidth]{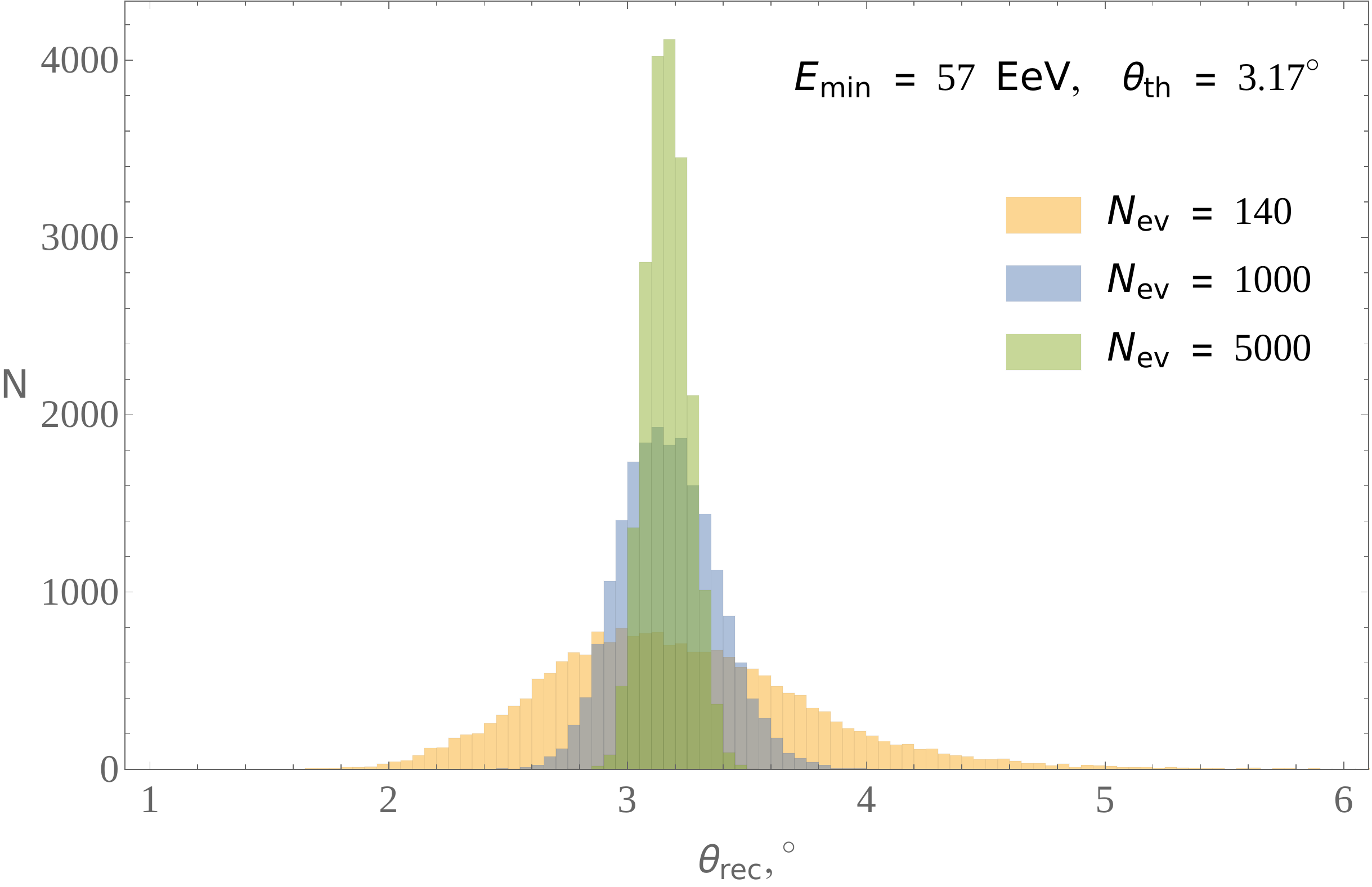}
\caption{
\label{fig:theta-theta}
Distribution of $TS$ minima, $\theta_{\rm rec}$, for event sets based on the same flux model map
($E_{\rm min} = 57$~EeV, $\theta_{\rm th} = 3.17^\circ$, no reg. GMF effect) but for different
number of events in a set: $N_{\rm ev} = 140$ (yellow histogram), $N_{\rm ev} = 1000$ (blue histogram) and
$N_{\rm ev} = 5000$ (green histogram).
}
\end{center}
\end{figure*}

We now 
construct the distribution of the $TS$ minima $\theta_{\rm rec}$ and compare its width with $\Delta \theta_{\rm rec}$. These distributions are shown in Fig.~\ref{fig:theta-theta} for several numbers of events $N_{\rm ev}$ in mock sets.
We found that already for $N_{\rm ev} \sim 100$ the deviation
of the width of $\theta_{\rm rec}$ distribution from the
mean $\Delta\theta_{\rm rec}$ is less then $3\%$, in agreement with the expectation from the  $\chi^2$-distribution.
One can see in Fig.~\ref{fig:theta-theta} that the distribution of $\theta_{\rm rec}$ for $N_{\rm ev}=140$ is slightly asymmetric and its maximum
is shifted from the input value $\theta_{\rm th} = 3.17^\circ$ to smaller values. 
However, as $N_{\rm ev}$ increases the distribution becomes more narrow and symmetric, and the accuracy of reconstruction of  $\theta_{\rm th}$ from this distribution increases. 

Having checked that we can recover the input flux smearing parameter in the absence of the regular magnetic field, let us see how the test statistics (\ref{eq:TS}) behaves when such a field is present. For this test we take the regular GMF model of Ref.~\cite{Pshirkov:2011um} and fix its overall magnitude in such a way that the mean deflection in the regular field is 3 times larger than the mean random deflection which we keep the same as in the beginning of this Section. Note that this is a realistic ratio between the two contributions \cite{Pshirkov:2013wka}, but regular deflections themselves are about 3 times larger than would be for protons and best-fit GMF parameters of Ref.~\cite{Pshirkov:2011um}. 

\begin{figure*}
\begin{center}
 \includegraphics[width=0.49\columnwidth]{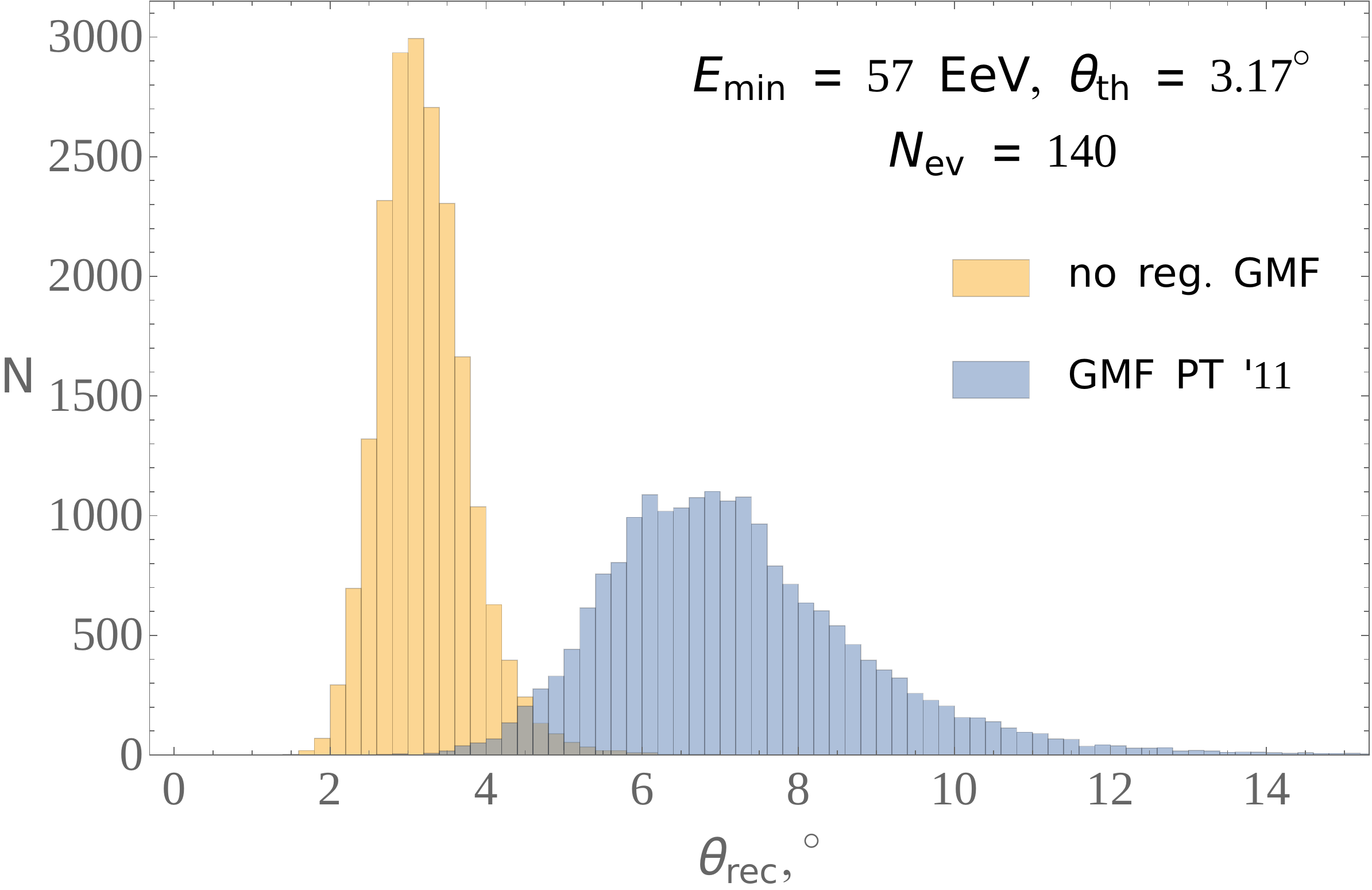}
 \includegraphics[width=0.49\columnwidth]{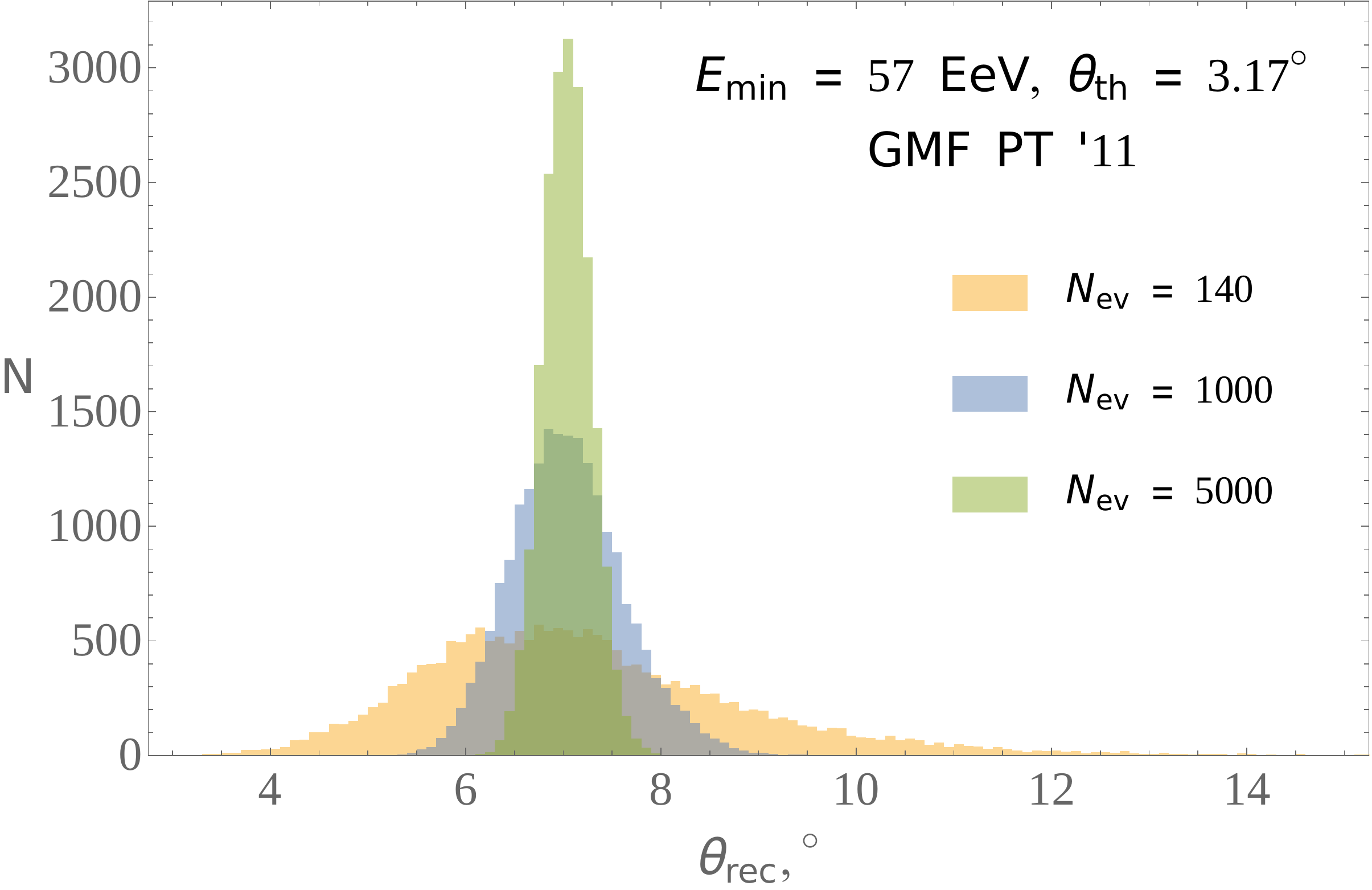}
\caption{
\label{theta_Q-theta}
{\it Left:} Comparison of $\theta_{\rm rec}$ distribution for event sets based on the flux model maps
with regular GMF effect (blue histogram) and without it (yellow histogram). Other event sets parameters
are the same for both sets: $E_{\rm min} = 57$~EeV, $\theta_{\rm th} = 3.17^\circ$ and $N_{\rm ev} = 140$.
{\it Right:} Distributions of $\theta_{\rm rec}$ for event sets with regular GMF effect for different
number of events in a set: $N_{\rm ev} = 140$ (yellow histogram), $N_{\rm ev} = 1000$ (blue histogram) and
$N_{\rm ev} = 5000$ (green histogram). Other event set parameters are the same as for left picture.
}
\end{center}
\end{figure*}
The results of the test are presented in Fig.~\ref{theta_Q-theta} where on the left panel we show the comparison with the case of zero regular field, and on the right panel the behavior of the distribution with the number of events in the mock set, $N_{\rm ev}$. Notably, individual  TS curves still have minima so that $\theta_{\rm rec}$ can be determined, and their distribution has a clear maximum, although shifted to larger angles. As in the case of no regular field, the resulting distributions tend to Gaussian ones as $N_{\rm ev}$ increases.

\section{Results}
\label{sec:results}
\subsection{Inference of the UHECR mass composition from the likelihood shape}
\label{sec:composition}
\begin{figure*}
\begin{center}
	\includegraphics[width=0.49\columnwidth]{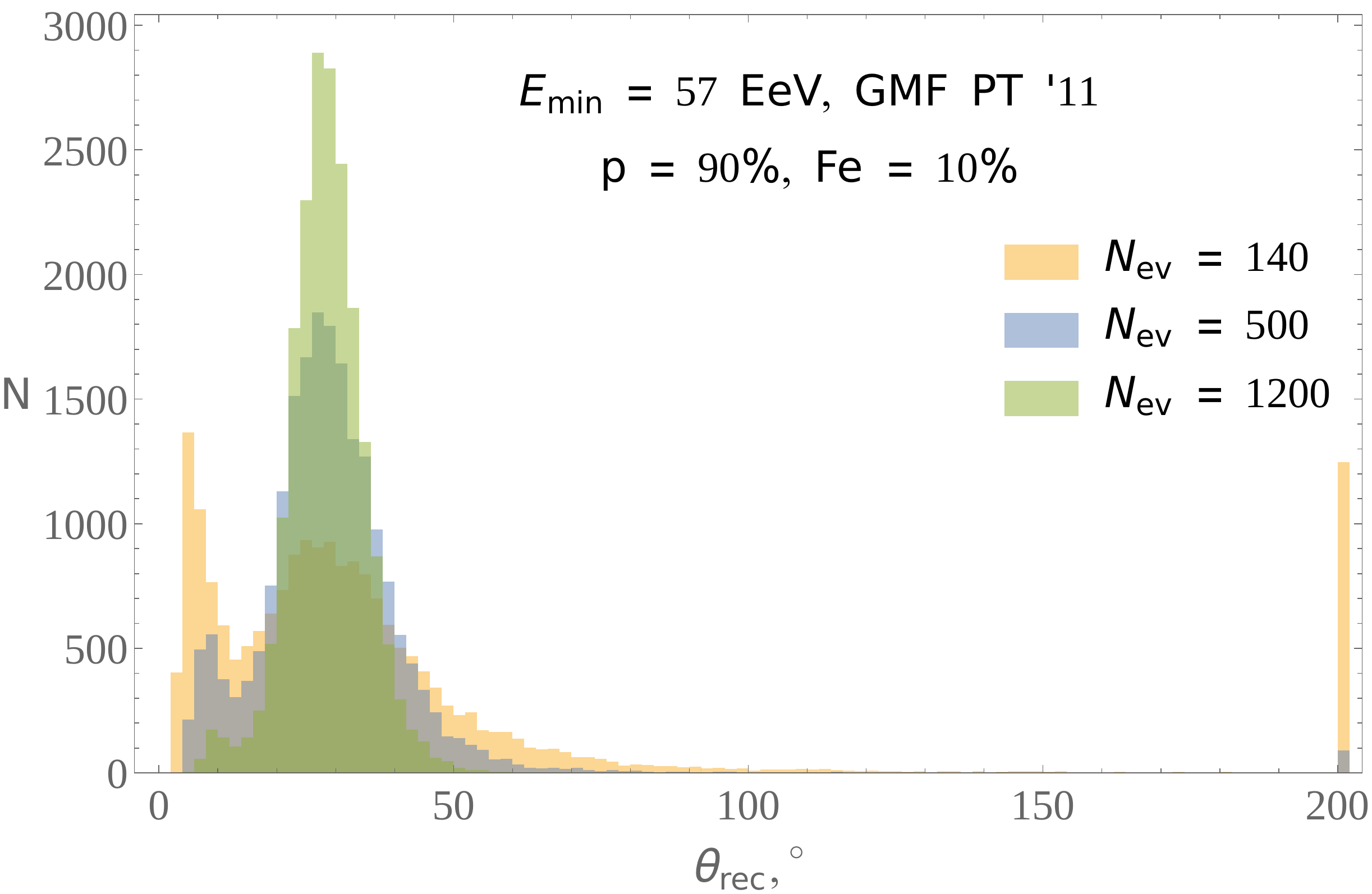}
	\includegraphics[width=0.49\columnwidth]{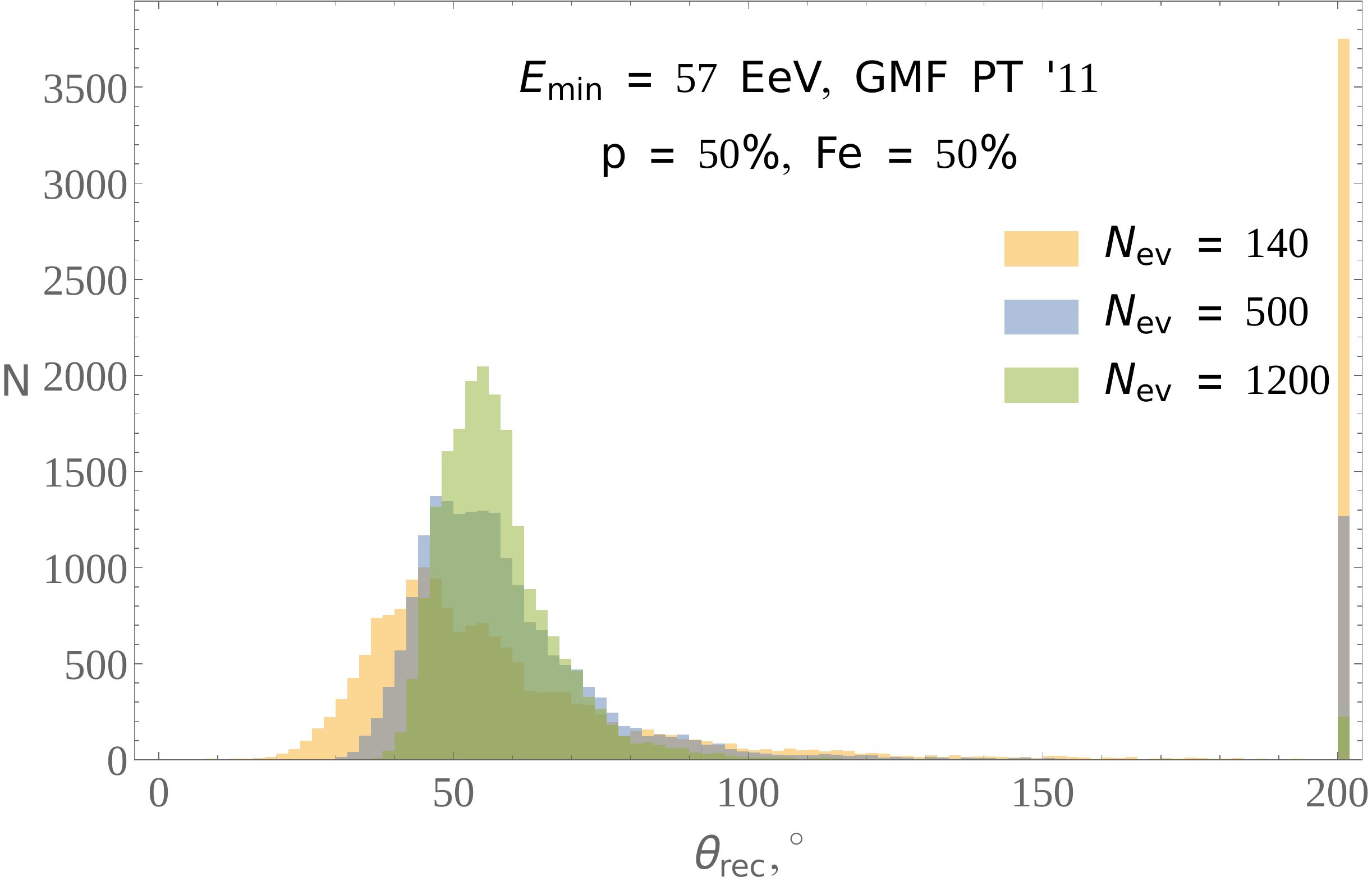}
	\includegraphics[width=0.49\columnwidth]{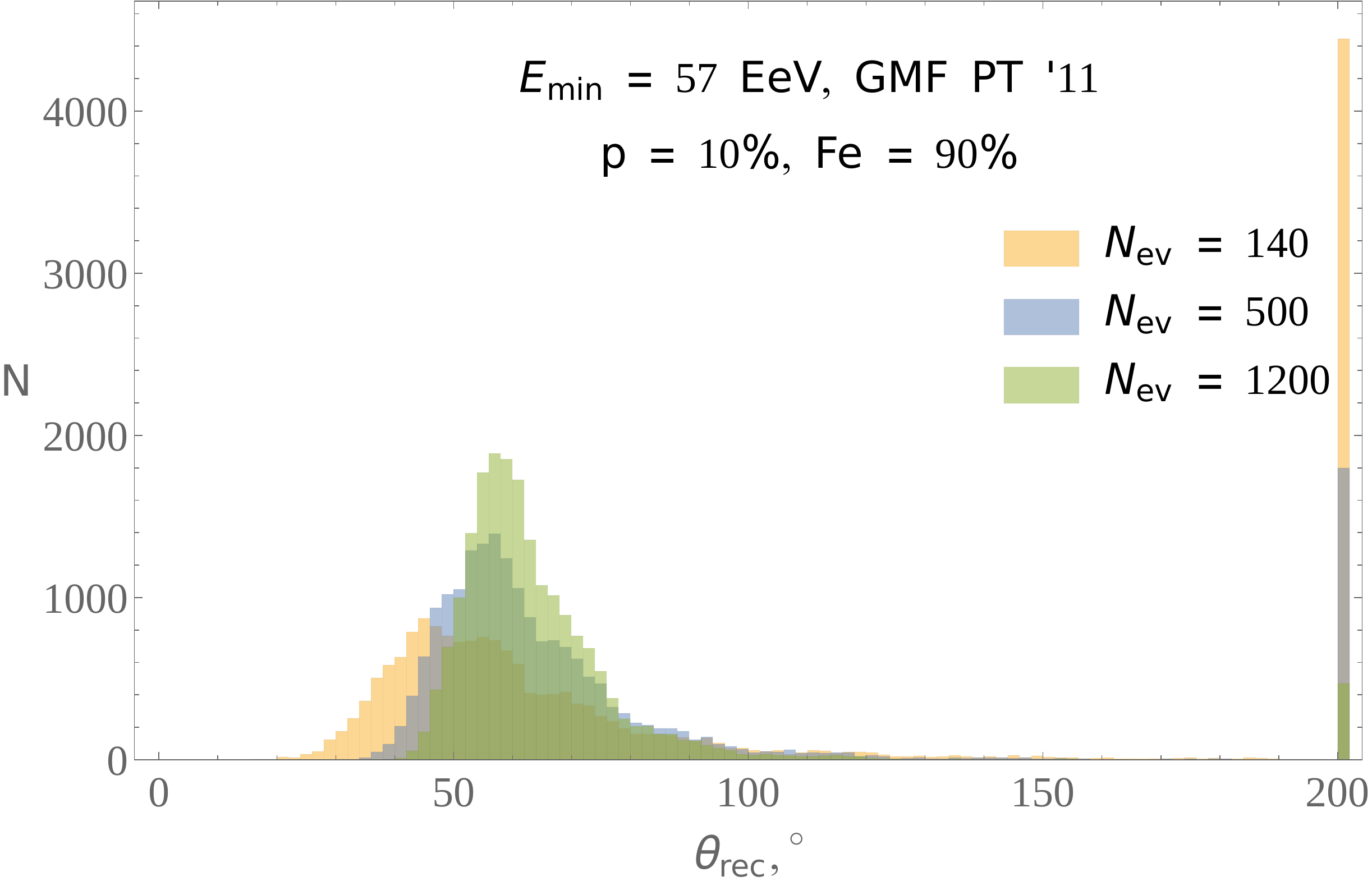}
\caption{
\label{fig:var_p-Fe}
Distributions of $\theta_{\rm rec}$ for event sets with $E_{\rm min} = 57$~EeV,
different constant p-Fe mix compositions
and various numbers of events in a set: $P_{\rm p} = 90\%, P_{\rm Fe} = 10\%$ (top left),
$P_{\rm p} = 50\%, P_{\rm Fe} = 50\%$ (top right) and $P_{\rm p} = 10\%, P_{\rm Fe} = 90\%$ (bottom).
Numbers of events are $N_{\rm ev} = 140$ (yellow histogram), $N_{\rm ev} = 500$ (blue histogram)
and $N_{\rm ev} = 1200$ (green histogram).
}
\end{center}
\end{figure*}

The robustness of the likelihood shape to the presence of regular GMF opens a way for a new method of
UHECR mass composition estimation. As it was shown, the position of the likelihood minimum, $\theta_{\rm rec}$, is the proxy of the primary particles deflection from their sources that is directly related to the
charge of these particles. Therefore, by measuring $\theta_{\rm rec}$ in the data one could estimate
the mean charge  of UHCER in a given sample.
This estimation can be made by comparing the value of  $\theta_{\rm rec}$ for the data with 
the distribution of $\theta_{\rm rec}$ in mock event sets of a particular UHECR composition model.
Since we only have one number $\theta_{\rm rec}$ determined from the data, the exact composition is impossible to determine  because the same value of $\theta_{\rm rec}$ may correspond to different composition models. Nevertheless, the composition can be constrained by excluding models where the
measured value of $\theta_{\rm rec}$ never occurs or occurs rarely.

To illustrate our method suppose a hypothetical experiment (we use the parameters of TA for concreteness) has observed $N_{\rm ev}$ events, calculated the test statistics (\ref{eq:TS}) and found its minimum to be $\theta_{\rm data}$. What conclusions regarding the composition of UHECR can be deduced from that?

As already explained, in this paper we limit ourselves with a simplified approach where the UHECR consist of a proton-iron mixture. The aim is thus to constrain the fraction of protons and iron in this mix. Despite the simplification, the results of this approach may still be of practical importance as the upper bound on the proton fraction derived in this setup is conservative in the sense that it will hold if iron is replaced by lighter species, because the iron component drags the maximum of the $\theta_{\rm rec}$ distribution to larger values stronger than any other possible admixture. The same applies to the upper bound on the fraction of iron --- the proton component pulls the maximum of $\theta_{\rm rec}$ to smaller values stronger than other nuclei. 

We will present the results for different number of events $N_{\rm ev}$ roughly corresponding to the UHECR statistics already accumulated and expected in the future. 
The number of events with $E > 57$~EeV accumulated to date by the surface detector of the TA experiment is $\sim 140$, 
while with a recently constructed extension, TAx4~\cite{Kido:2020isy}, one expects the tripling of this statistics
in the next six years. At corresponding energies, the current statistics accumulated by the Pierre Auger surface detector is about $\sim 1200$ events~\cite{diMatteo:2020dlo}.  Therefore, we present upper limits for the proton and iron fractions for $N_{\rm ev}$ equal to 140, 500 and 1200. 

For each $N_{\rm ev}$ we generate 20000 mock event sets with energy-independent 
proton and iron fractions $P_{\rm p}$ and $P_{\rm Fe}= 1-P_{\rm p}$, respectively. We repeat this
procedure for various values of $P_{\rm p}$ and $P_{\rm Fe}$. The proton and iron events are generated from flux maps computed with corresponding attenuation functions. These maps are processed through the same GMF of Ref.~\cite{Pshirkov:2011um} with the best fit parameters and charges 1 and 26 for proton and iron, respectively. They are then smeared with the latitude-dependent Gaussian width defined by Eq.~(\ref{eq:smearing}) for protons at $E=40$~EeV and rescaled according to the energy of the current bin and particle charge. No free parameters enter this calculation apart from the fractions $P_{\rm p}$ and $P_{\rm Fe}$. For every value of $P_{\rm p}$ we build a distribution of the minima of our test statistics,  $\theta_{\rm rec}$. The illustrative examples of these distributions for three different values of $P_{\rm p}$  are shown in Fig.~\ref{fig:var_p-Fe}. 

\begin{figure*}
\begin{center}
	\includegraphics[width=0.49\columnwidth]{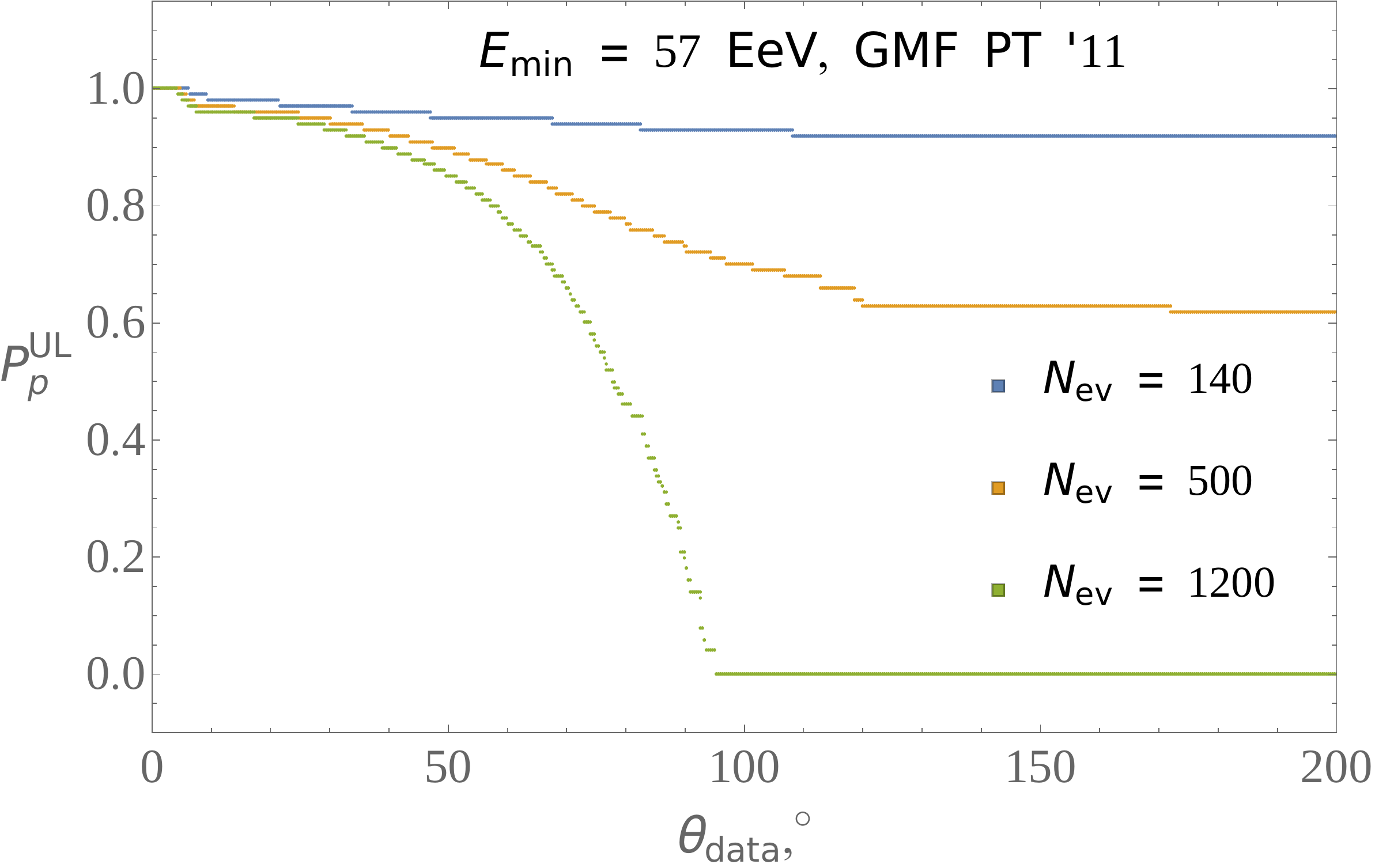}
	\includegraphics[width=0.49\columnwidth]{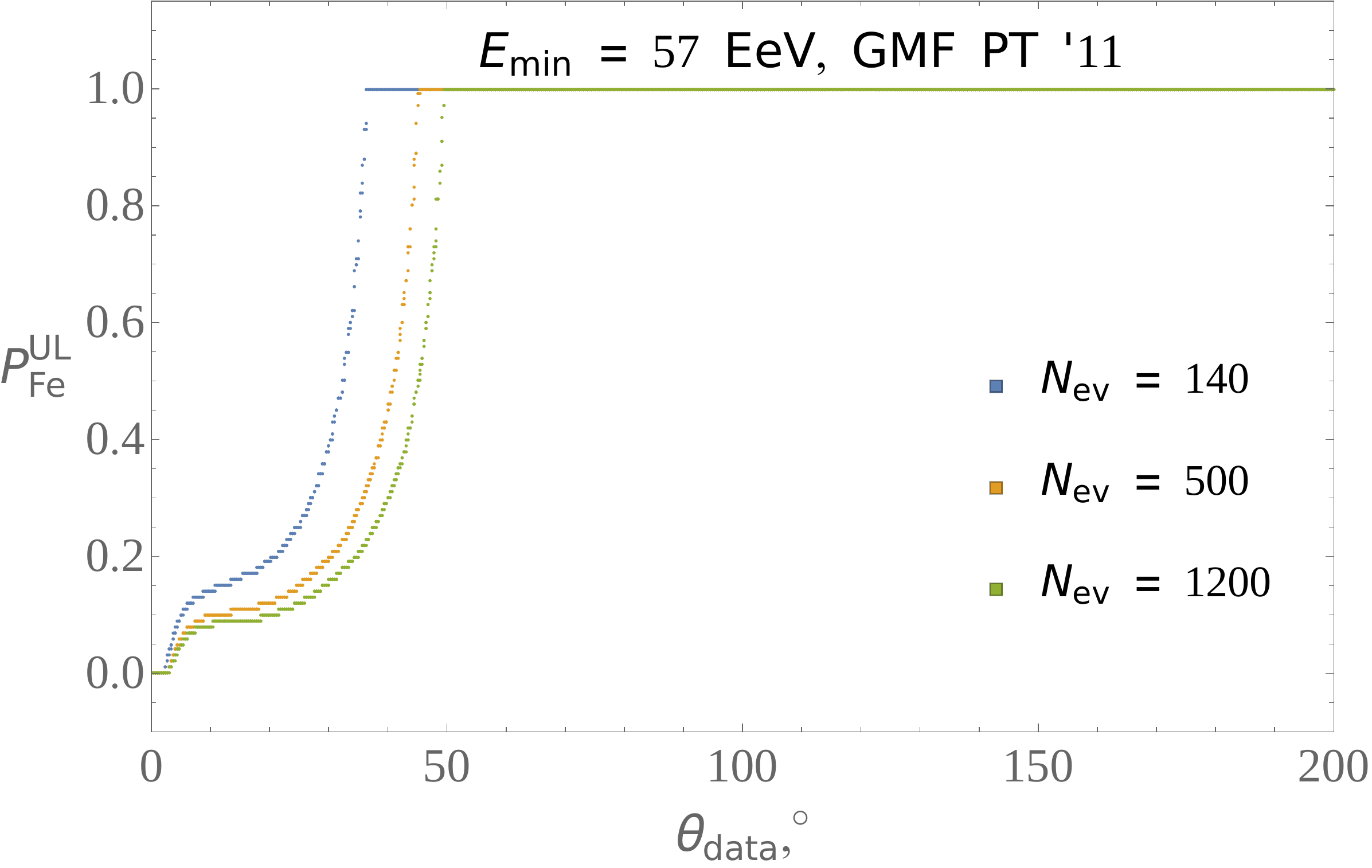}
\caption{
\label{fig:UL_p-Fe}
$95\%$~C.L. upper-limits on the constant fractions of proton (left panel) and
iron (right panel) in p-Fe mix as a function of $\theta_{\rm data}$.
Numbers of events are $N_{\rm ev} = 140$ (blue line), $N_{\rm ev} = 500$ (yellow line)
and $N_{\rm ev} = 1200$ (green line). $E_{\rm min} = 57$~EeV.
}
\end{center}
\end{figure*}

Given the assumed value of $\theta_{\rm data}$, we identify values of $P_p$ such that either the right tail of the histogram $\theta_{\rm rec}>\theta_{\rm data}$ or the left tail $\theta_{\rm rec}<\theta_{\rm data}$ contains no more than 5\% of occurrences. In the first case we conclude that the proton fraction $P_p$ and larger ones are excluded at 95\%~C.L., while in the second that iron fraction $P_{\rm Fe}= 1-P_{\rm p}$ and larger ones are excluded at 95\%~C.L. 
The resulting constraints are plotted as a function of $\theta_{\rm data}$ in Fig.~\ref{fig:UL_p-Fe}
We also summarize several numerical values of upper limits derived for three sample values of $\theta_{\rm data}$ in Table~\ref{tab:limits_fixed}. 

As expected, when $\theta_{\rm data}$ is large one cannot constrain the
fraction of iron nuclei, while at small $\theta_{\rm data}$ no constraint on
the fraction of protons can be set. Less obvious is that constraints on the
fraction of iron are generally stronger than those on the fraction of protons, as is clear from the fact that the left tail of the $\theta_{\rm rec}$-distributions is steeper than the right one. The underlying reason is that an admixture of protons on a low-contrast mostly iron map has smaller effect on the test statistics than an admixture of iron on a higher-contrast mostly proton map. 

\begin{table*}
\begin{center}
\begin{tabular}{|c|c|c|c|}
\hline
$\theta_{\rm rec}$ & $N_{\rm ev}$ & $P_{\rm p}^{\rm UL}$ ($95\%$ C.L.) & $P_{\rm Fe}^{\rm UL}$ ($95\%$ C.L.) \\
\hline
$3.0^\circ$ & 140  & 1.00 & 0.04 \\ \hline
$3.0^\circ$ & 500  & 1.00 & 0.01 \\ \hline
$3.0^\circ$ & 1200 & 1.00 & 0.00 \\ \hline
\hline
$40^\circ$  & 140  & 0.96 & 1.00 \\ \hline
$40^\circ$  & 500  & 0.93 & 0.45 \\ \hline
$40^\circ$  & 1200 & 0.90 & 0.30 \\ \hline
\hline
$200^\circ$ & 140  & 0.92 & 1.00 \\ \hline
$200^\circ$ & 500  & 0.62 & 1.00 \\ \hline
$200^\circ$ & 1200 & 0.00 & 1.00 \\ \hline
\end{tabular}
\end{center}
\caption{
$95\%$~C.L. upper-limits on fractions of protons, $P_{\rm p}$, and fraction of iron nuclei, $P_{\rm Fe}$,
in p-Fe mix models with $P_{\rm p}$ and $P_{\rm Fe}$ not changing with energy.
The limits are derived from the mock sets with various number of events and
different values of TS minimum: $\theta_{\rm rec} = 3.0^\circ$ (top panel),
$\theta_{\rm rec} = 40^\circ$ (middle panel) and isotropic set ($\theta_{\rm rec} = 200^\circ$, bottom panel).
}
\label{tab:limits_fixed}
\end{table*}

So far we have been assuming that the composition does not change with energy. However, since the test statistics (\ref{eq:TS}) depends differently on composition at different energies, our method may also be used to distinguish a composition evolving with energy from the constant composition. To illustrate this point we compare distributions of $\theta_{\rm rec}$
for several composition models: pure proton constant composition ($M_1$),
proton-iron mix with constant $P_{\rm p} = 0.9$ ($M_2$), proton-iron mix with $P_{\rm p} = 0.9 \cdot (57~{\rm EeV} / E)^2$ ($M_3$) and the main composition model from
Ref.~\cite{Aab:2016zth} --- a mix of nitrogen and silicon
($M_4$).~\footnote{For the purpose of this illustration, we generate the flux maps for nitrogen and silicon nuclei assuming proton attenuation length which is larger than the actual one. The resulting maps are more proton-like (less contrast) than they should actually be,  and our model comparison is therefore conservative --- with the correct attenuation the models will be easier to distinguish.}
We compare these models in pairs by choosing one model $M_{\rm test}$ as a ``null hypothesis'' (the one to be constrained, or test model) and another, $M_{\rm ref}$, as an ``alternative hypothesis'' (the reference model the data are assumed to follow). Following the standard definition of the statistical power  we cut out  5\%-tail of the null hypothesis distribution on the side where it overlaps most with the alternative one, and integrate the alternative distribution from the cut point to infinity to get the statistical power $P(M_{\rm ref}, M_{\rm test})$. 
The resulting value, $P(M_{\rm ref}, M_{\rm test})$, is interpreted as
a chance to constrain the test model at $95\%$~C.L. if the data follows the reference model. 
When two distributions overlap only slightly, the statistical power is close to 1. 

\begin{figure*}
\begin{center}
\includegraphics[width=0.49\columnwidth]{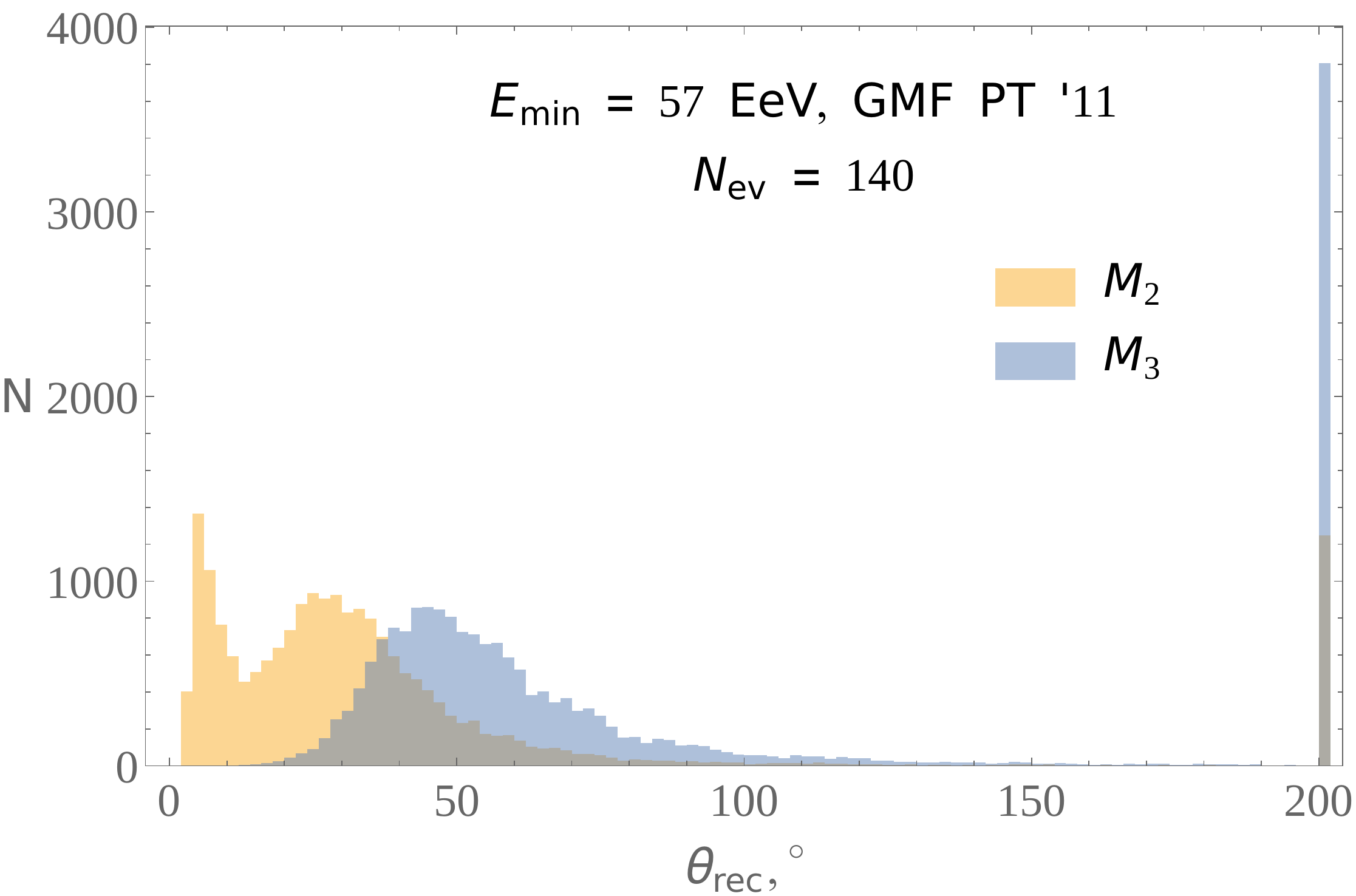}
\includegraphics[width=0.49\columnwidth]{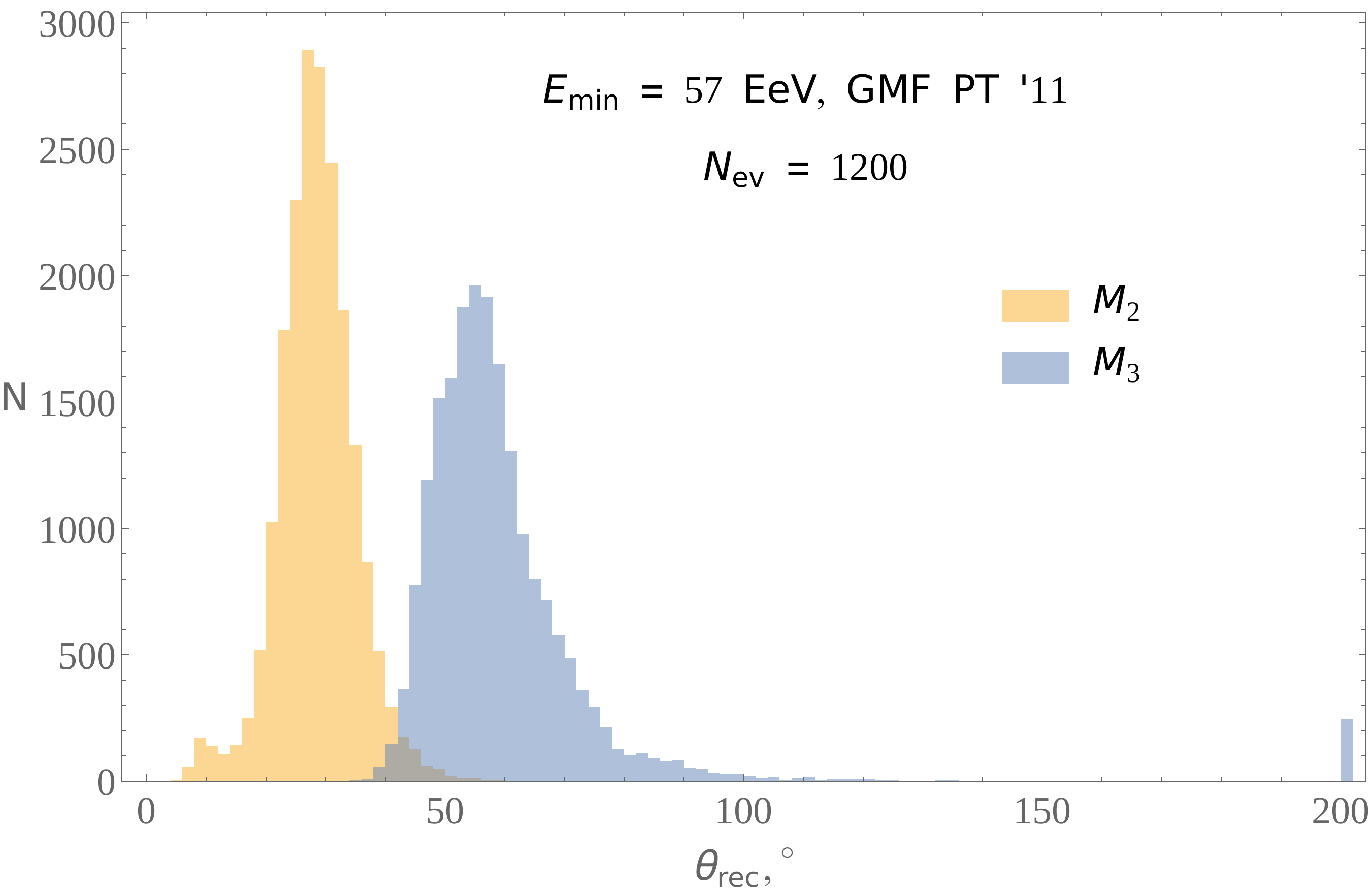}
\includegraphics[width=0.49\columnwidth]{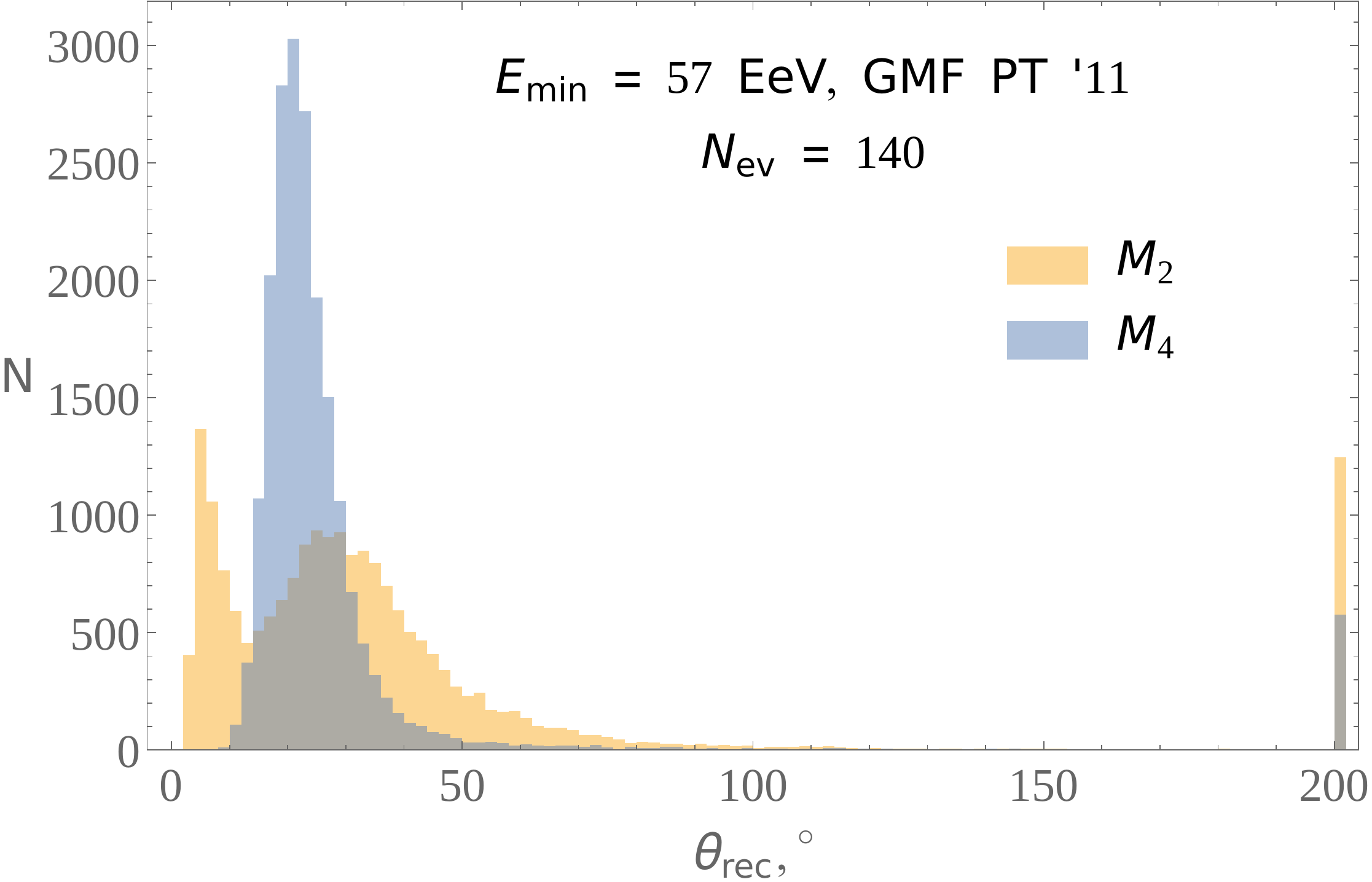}
\includegraphics[width=0.49\columnwidth]{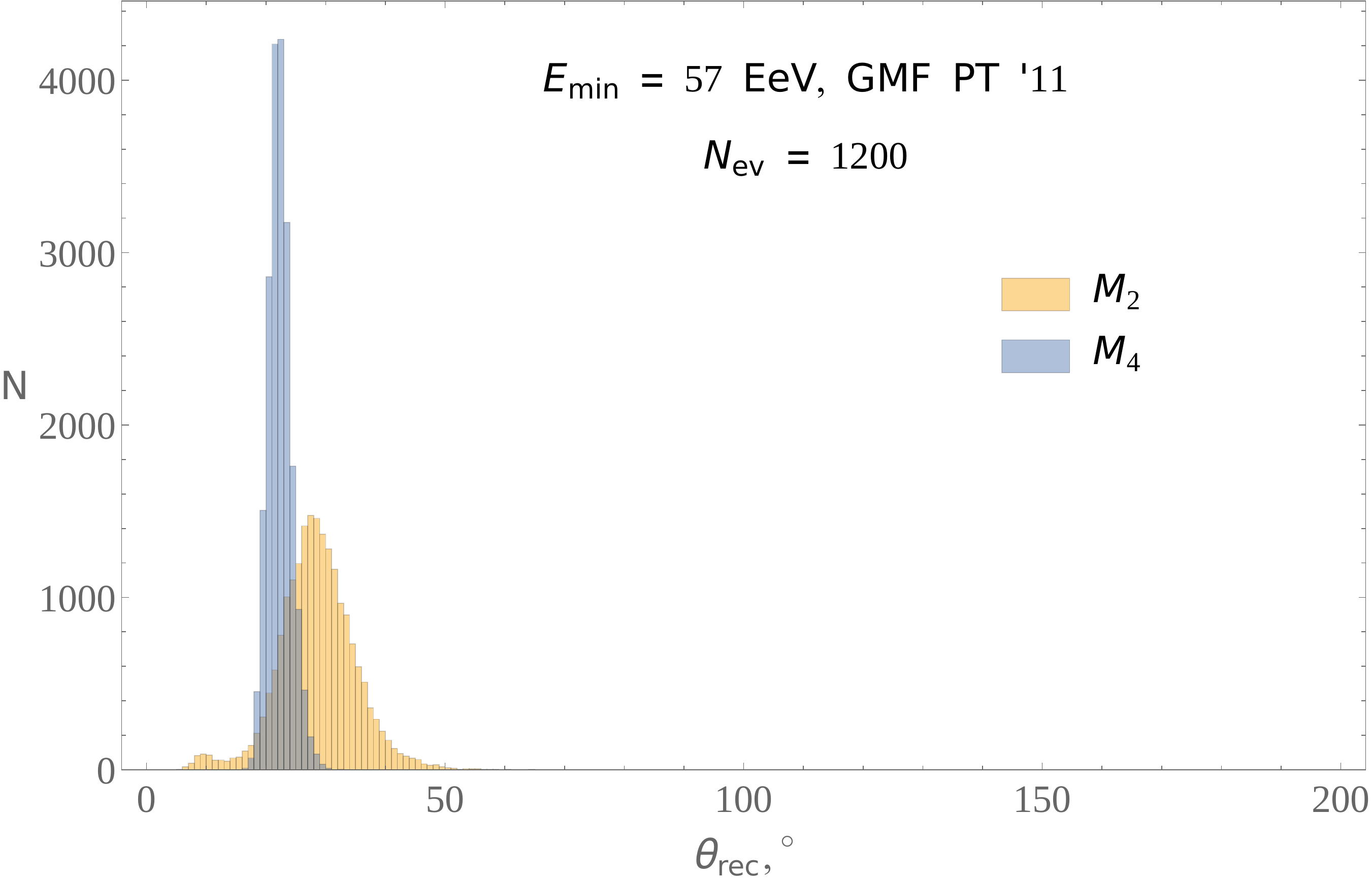}
\caption{
\label{fig:change_comp_dist_compar}
Examples of $\theta_{\rm rec}$ distributions overlap between $M_2$ and $M_3$ composition models
for $N_{\rm ev} = 140$ (top left) and $N_{\rm ev} = 1200$ (top right) and between $M_2$ and $M_4$ composition models for $N_{\rm ev} = 140$ (bottom left) and $N_{\rm ev} = 1200$ (bottom right). See text for models description.
}
\end{center}
\end{figure*}

The comparison of the models for various values of $N_{\rm ev}$ is shown in Fig.~\ref{fig:change_comp_dist_compar}.
The respective values of $P(M_{\rm ref}, M_{\rm test})$ are given in Tab.~\ref{tab:limits_overlap}.
One can see that at large enough statistics our method allows one to distinguish pure proton composition from a proton-dominated one, as well as proton dominated composition from a
composition that becomes heavier with energy.
However, even a small admixture of iron leads to significant decrease
in the separation power (cf. $P(M_1, M_3)$ and $P(M_2, M_3)$) which, however, grows with statistics.
It is also worth noting that one can certainly tell pure proton model $M_1$ from the medium-mass nuclei mix $M_4$, and the proton-iron mix $M_2$ from the medium-mass mix $M_4$ --- with the reasonable confidence. 

\begin{table*}
\begin{center}
\begin{tabular}{|c|c|c|c|c|c|c|}
\hline
$N_{\rm ev}$ & $P(M_1, M_2)$ & $P(M_1, M_3)$ & $P(M_2, M_3)$ & $P(M_1, M_4)$ & $P(M_2, M_4)$ \\
\hline
140  & 0.95 & 1.00 & 0.58 & 1.00 & 0.25 \\ \hline
500  & 1.00 & 1.00 & 0.89 & 1.00 & 0.50 \\ \hline
1200 & 1.00 & 1.00 & 0.99 & 1.00 & 0.69 \\ \hline
\end{tabular}
\end{center}
\caption{
Values of statistical power to constrain a test composition model at $95\%$~C.L. if data follows a reference model, see text for explanation.
Results for various number of events in a set are presented.
}
\label{tab:limits_overlap}
\end{table*}

\subsection{Uncertainties}
\label{sec:uncertainties}

There are several possible sources of uncertainties in our method of assessing the cosmic ray composition: source distribution, injected composition, extragalactic and galactic magnetic field properties. As it was discussed in Sect.~\ref{sec:model} our strategy is to minimize these uncertainties by using well motivated observational and theoretical assumptions whenever possible. In this Section we discuss the robustness of our assumptions and estimate the impact of unavoidable uncertainties on the performance of the method.

As it was discussed in Section~\ref{sec:propagation}, for the purpose of performance estimation we have assumed a proton-iron injection mix and treat its propagation neglecting the secondary species. Apart from a technical simplification, these assumptions make the constraints on the composition more conservative. Indeed, for a given proton fraction $P_{\rm p}$ the deflections are largest when the remaining part is pure iron. 
Likewise, for a given iron fraction $P_{\rm Fe}$ the deflection of p-Fe is smaller than that for any multi-component mix. This implies that a more realistic treatment of injection composition and/or account for secondaries would make the resulting constraints on the corresponding fraction only stronger.

The main uncertainty of our method comes from magnetic fields. It can be divided into four independent parts: the uncertainty of the regular GMF structure, of the overall regular GMF strength, of the random GMF strength, and the uncertainty of the EGMF strength and structure. The latter three uncertainties are characterized by one parameter each. The random GMF can be parameterized by $\theta_{\rm th}$  defined as the smearing angle at the Galactic pole for protons with energy $E=100$~EeV (recall that we adopted latitude-dependent random deflections given by Eq.~(\ref{eq:smearing})) and rescaled according to particle charge and energy. So far we kept  $\theta_{\rm th}=0.4^\circ$ fixed, but now we will vary this parameter. The overall magnitude of the regular GMF can be parameterized by the dimensionless normalization factor $Q$ with the value $Q=1$ for the best-fit parameters used up to now. Finally, the deflections in EGMF enter as an additional uniform smearing parameter $\theta_{\rm EGMF}$ (it has been set to zero in our reference flux model).

\begin{figure*}
\begin{center}
 \includegraphics[width=0.49\columnwidth]{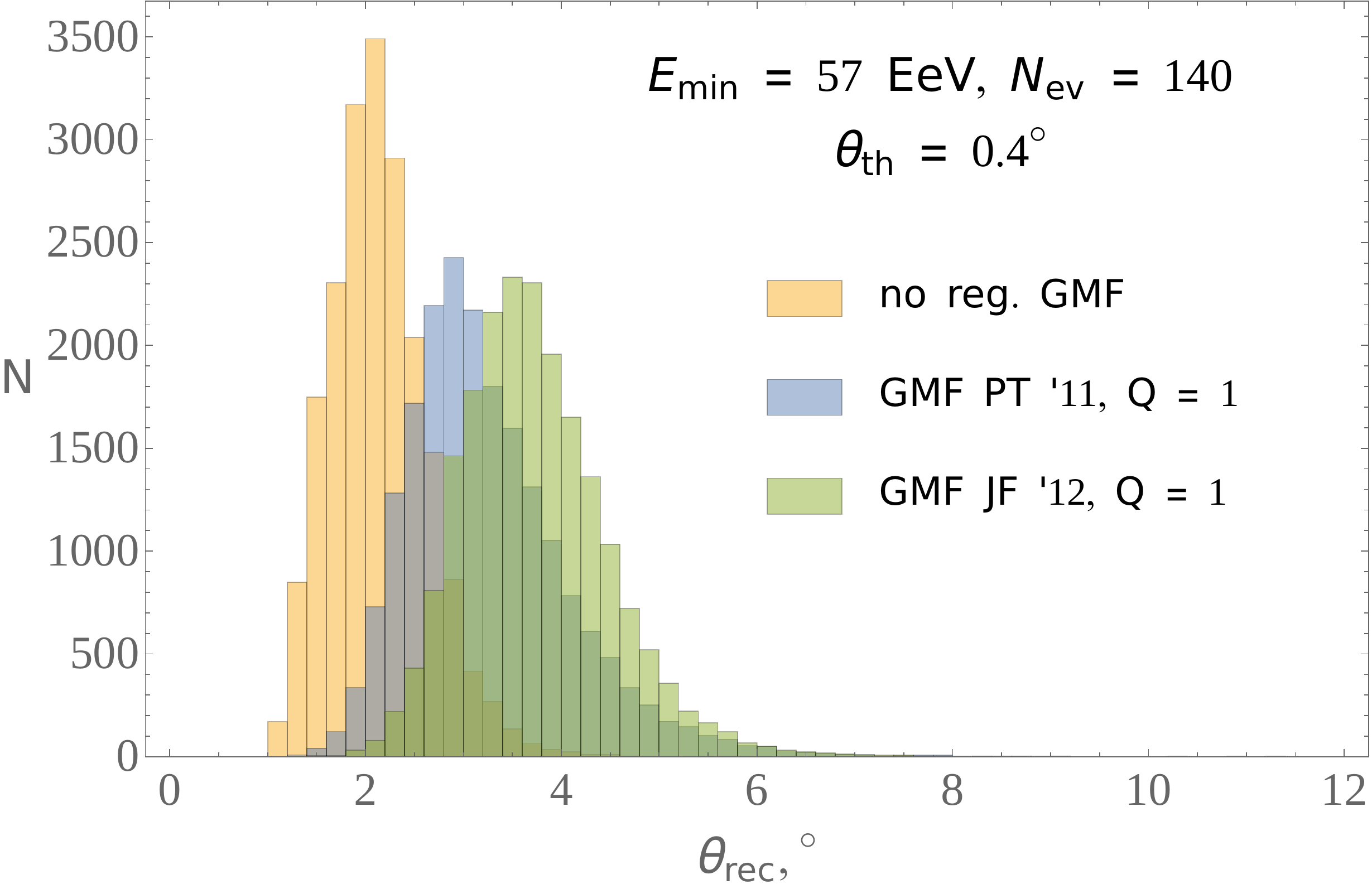}
 \includegraphics[width=0.49\columnwidth]{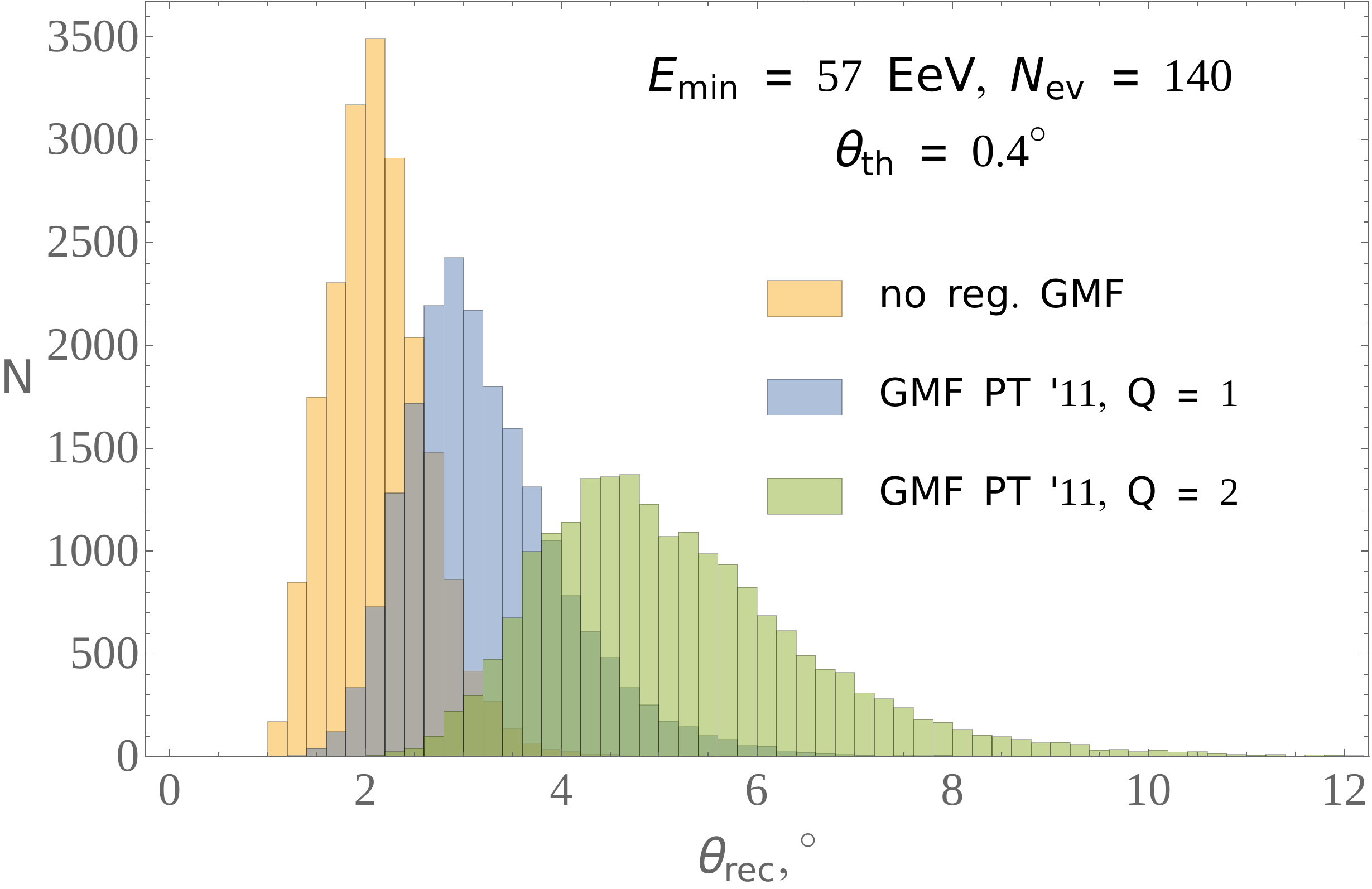}
 \includegraphics[width=0.49\columnwidth]{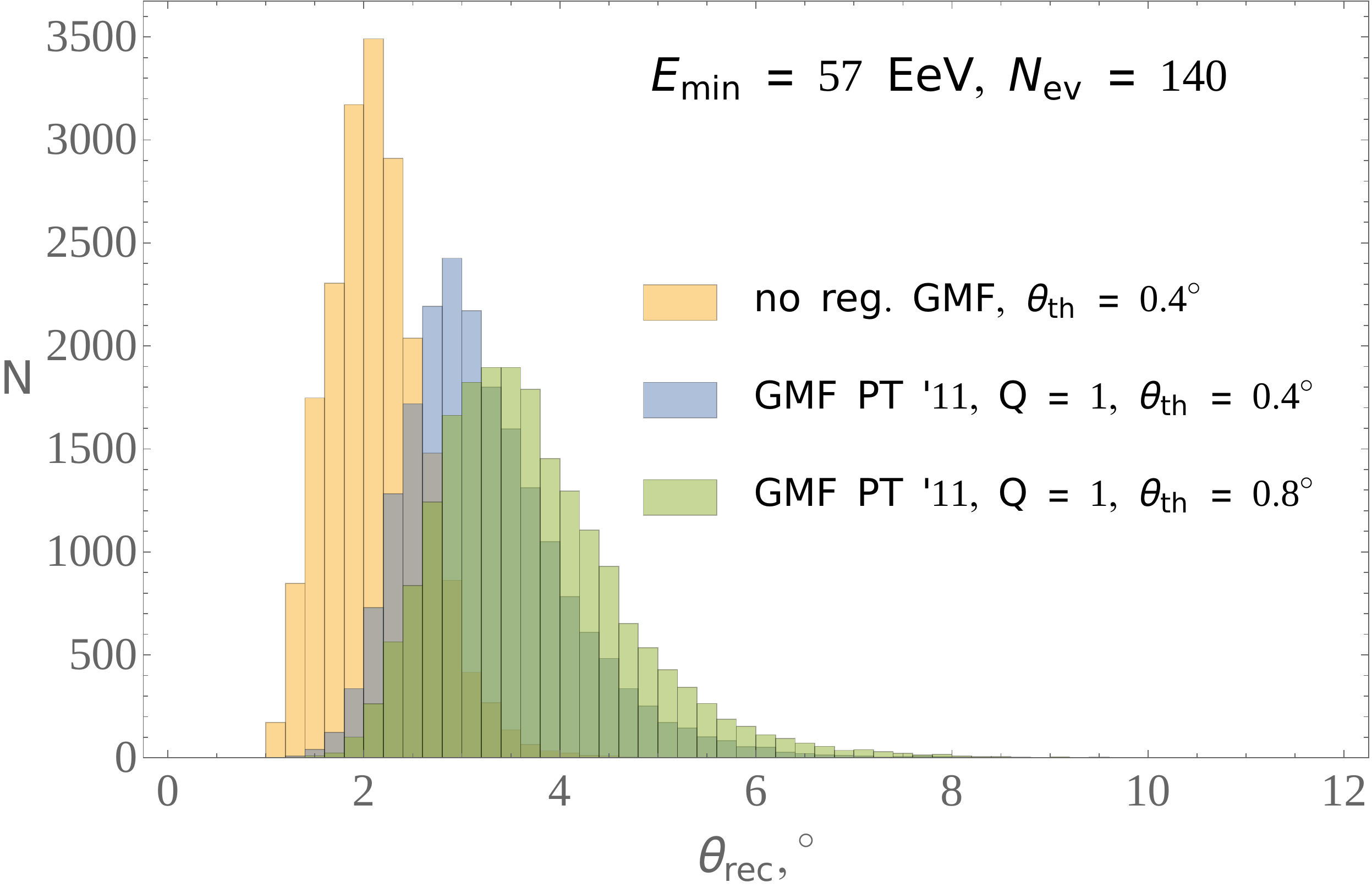}
 \includegraphics[width=0.49\columnwidth]{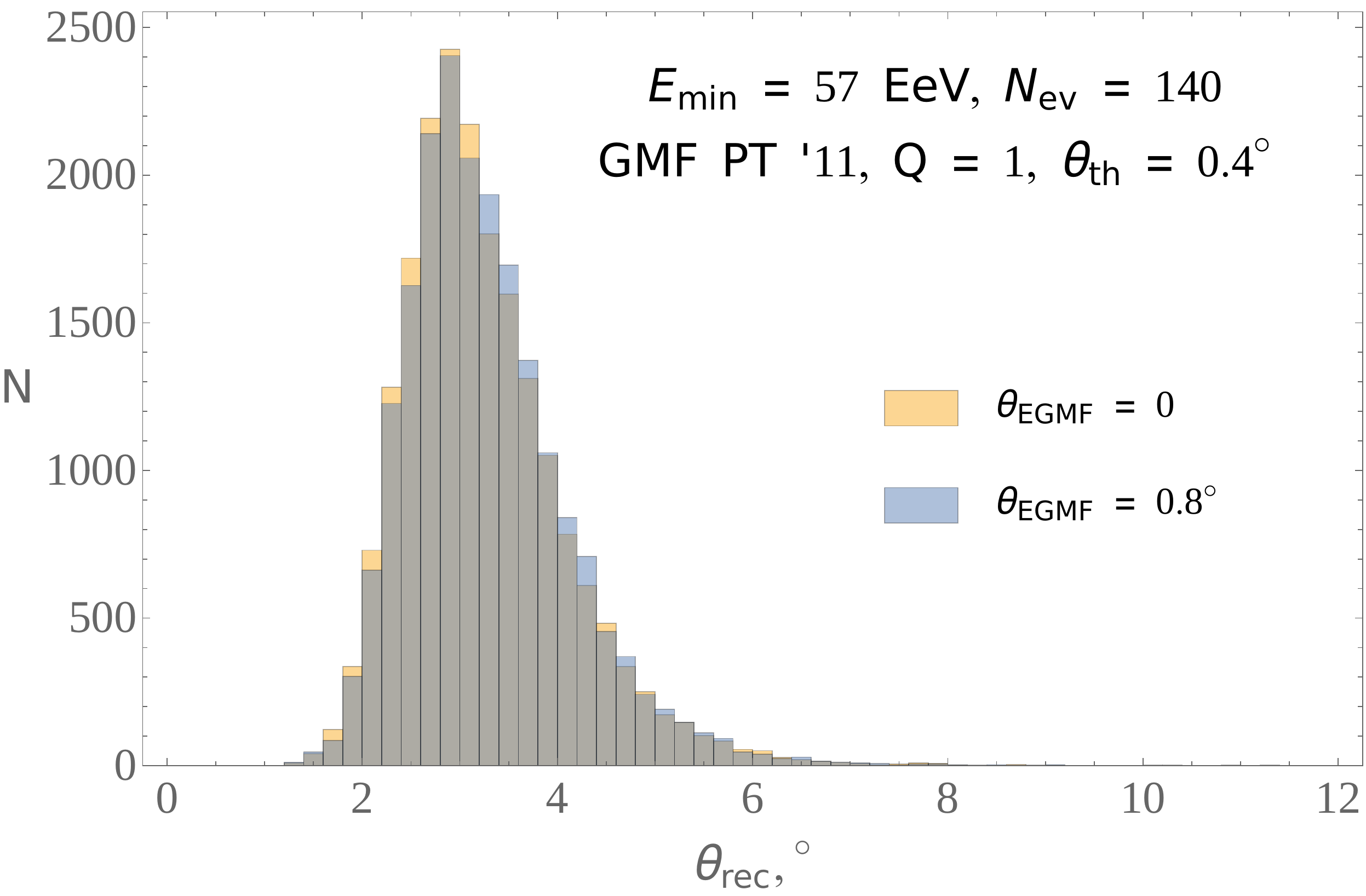}
\caption{
\label{fig:theta_dist_uncert}
Comparison of $\theta_{\rm rec}$ distributions for various values of GMF parameters.
{\it Top left:} no regular GMF vs. regular GMF model of Ref.~\cite{Pshirkov:2011um} and model
of Ref.~\cite{Jansson:2012pc}, for both reg. GMF models $Q = 1$ and $\theta_{\rm th} = 0.4^\circ$.
{\it Top right:} regular GMF model of Ref.~\cite{Pshirkov:2011um}, $\theta_{\rm th} = 0.4^\circ$, comparison of $Q = 1$ and $Q = 2$.
{\it Bottom left:} regular GMF model of Ref.~\cite{Pshirkov:2011um}, $Q = 1$, comparison of $\theta_{\rm th} = 0.4^\circ$ and $\theta_{\rm th} = 0.8^\circ$.
{\it Bottom right:} regular GMF model of Ref.~\cite{Pshirkov:2011um}, $Q = 1$, $\theta_{\rm th} = 0.4^\circ$, comparison of $\theta_{\rm EGMF} = 0$ and $\theta_{\rm EGMF} = 0.8^\circ$.
Pure proton composition with $E_{\rm min} = 57$~Eev and $N_{\rm ev} = 140$ are imposed for all event sets. 
}
\end{center}
\end{figure*}

In Fig.~\ref{fig:theta_dist_uncert} we vary these parameters from their reference values one at a time and show how this affects the distribution of $\theta_{\rm rec}$ for the pure proton composition model and the GMF of Ref.~\cite{Pshirkov:2011um}. We also show the comparison of the two GMF models of Refs.~\cite{Pshirkov:2011um, Jansson:2012pc}. The impact of uncertainties on the composition constraints is encoded in the change of the histogram shape and position. One can see that changing the regular field normalization by a factor 2 has a much larger effect than switching between the two GMF models, supporting the statement that the method is not sensitive to the structure of the regular galactic field.

The impact of these uncertainties on the model comparison can be characterized by a relative difference $\varepsilon$ between statistical powers calculated with reference and test magnetic field parameters. For instance, in case of $P(M2, M3)$ we define
\be
\varepsilon = \frac{P_{\rm MF_{\rm test}}(M_2, M_3) - P_{\rm MF_{\rm ref}}(M_2, M_3)}{P_{\rm MF_{\rm ref}}(M_2, M_3)}\;.
\ee
First we reproduce the previous results for the regular GMF model of Ref.~\cite{Jansson:2012pc}
as a test model, while keeping the absolute strength of random and regular GMF components
fixed at their reference values $\theta_{\rm th} = 0.4^\circ$, $Q = 1$  and $\theta_{\rm EGMF} = 0$. 
When testing three other uncertainties we fix the regular GMF model of Ref.~\cite{Pshirkov:2011um} and change the parameters $\theta_{\rm th}$, $Q$ and  $\theta_{\rm EGMF}$  from their reference values one at a time. 
We consider the values $\theta_{\rm th} = 0.4^\circ, 0.8^\circ$, $Q = 1, 2$  and $\theta_{\rm EGMF} = 0, 0.8^\circ$.
The results are summarized in 
Table~\ref{tab:uncert}.

One can see that increase of GMF or EGMF magnitude or change of GMF model of Ref.~\cite{Pshirkov:2011um} to model of Ref.~\cite{Jansson:2012pc} indeed results in decrease in statistical power of the method at a given event statistics. 
In all cases, however, the impact of uncertainties decreases with increasing statistics, as expected from the definition of the likelihood \eqref{eq:TS}.

Thus, once the MF structure  and magnitude  are fixed, two different composition models yield peaks at
different positions, although more statistics is needed to distinguish the same composition models for a larger MF strength. The degeneracy between two composition models could occur
only in the situation when GMF or EGMF is so large that average deflections are larger then the fiend of view of the experiment (about a half of the sky for TA), in which case  
both models have peaks at the value of $\theta_{\rm rec}$ corresponding to isotropy.
Therefore, the proposed method has a potential to separate the composition models
irrespectively of which MF model is assumed, provided that the event set is large enough.


Finally, consider our basic assumption that the sources are sufficiently numerous and can be treated on statistical basis.
If the sources are too rare to populate nearby galaxy clusters, Eq.~(\ref{eq:TS}) in its present form (i.e., based on the LSS mass distribution) can still be used to define a test statistics. One would expect that the method will still work (likely with a lower sensitivity) because in this case the sources will still correlate with the concentrations of matter, while the test statistics (\ref{eq:TS}) is degenerate with respect to moving events within the region of the same flux intensity. It is important to note, however, that given a high isotropy of the UHECR data, the case of very few sources within the GZK volume is by itself constrained from the same data.

To illustrate this argument we reproduced our simulations generating events from a number of catalogs with rare sources. Namely, from our original catalog we cut a $250$~Mpc volume-limited sample and further reduce it randomly to $1/50$ of its original size. We repeat this procedure a number of times to produce a bunch of source catalogs  containing an average of $7$ sources in a sphere of $50$~Mpc each. This corresponds to source density of $\sim 10^{-5}$~Mpc$^{-3}$. We generate a large number of mock UHECR sets for each of these catalogs following our general procedure described in Section~\ref{sec:mock_sets}. To be conservative we assume a pure iron composition, i.e. maximum deflections. 

The deviation of a given set from isotropy is characterized by the depth of the minimum of the test statistics (\ref{eq:TS}). We calculated this minimum for each of the generated sets and built the distribution of these values. 
We found that in 53\% of realisations the isotropy is excluded at more than $2\sigma$ level. Given our very conservative assumption about iron composition, this shows that source densities much smaller than $10^{-5}$~Mpc$^{-3}$ may be excluded if the data are close to isotropy. Thus, we conclude that our TS may allow one to constrain either the UHECR composition or the UHECR source density from one and the same arrival direction distribution. We leave a systematic analysis of the latter opportunity for the future.

\begin{table*}
\begin{center}
\begin{tabular}{|c|c|c|c|c|}
\hline
$N_{\rm ev}$ & $\varepsilon$, $\Delta {\rm GMF\; model}$ & $\varepsilon$, $\Delta \theta_{\rm th}$ & $\varepsilon$, $\Delta Q$ &  $\varepsilon$, $\Delta \theta_{\rm EGMF}$  \\
\hline
140  & -0.25 & -0.16  & -0.34 &  -0.28   \\ \hline
500  & -0.23 & -0.13  & -0.28 &  -0.19   \\ \hline
1200 & -0.10 & -0.039 & -0.13 &  -0.063  \\ \hline
\end{tabular}
\end{center}
\caption{
Impact of variation of magnetic field parameters on the method's statistical power
between composition models $M_2$ and $M_3$ for various number of events in a set, see text for details.
}
\label{tab:uncert}
\end{table*}

\subsection{Dependence on the lower energy threshold}
\label{sec:low_energy}
So far we considered the energy range $E>57$~EeV, but the same method can be applied at lower energies. This will address a different physics question, namely what is the composition in that energy range. 
However, it is instructive to compare the performance of the method for one and the same composition model
but with different energy thresholds, $E_{\rm min}$. The UHECR statistics at lower energies increases, but the events deflections and their uncertainties grow as well.
Therefore, it is difficult to estimate the change in a method performance {\it a priori}.
We estimate the sensitivity using simulations for $E > 10$~EeV with the same energy binning of 10 bins per decade.
We fix the size of the sample to 5000 events, which approximately corresponds to the recent statistics
of TA SD at these energies, and also corresponds to 140 events
in a sample with $E > 57$~EeV that was studied
in the main part of this work. Like in Section~\ref{sec:composition} we calculate the statistical power which determines a
chance to discriminate one composition model from another. 
Specifically,  we consider 6 proton-iron mix models with power-law change of the proton
fraction $P_{\rm p}$ with energy:
three models with $P_{\rm p} = 0.75$ at $10$~EeV and different power-law indices $P_{\rm p} = {\rm const}$ ($M_5$),
$P_{\rm p} \sim E^{-0.1}$ ($M_6$) and $P_{\rm p} \sim E^{0.1}$ ($M_7$), 
and three models with $P_{\rm p} = 0.5$ at $10$~EeV, also with different power-law behaviours $P_{\rm p} = {\rm const}$ ($M_8$), $P_{\rm p} \sim E^{-0.5}$ ($M_9$) and $P_{\rm p} \sim E^{0.5}$ ($M_{10}$).
We choose the model $M_5$ as the reference model for the models $M_6$ and $M_7$, and
the model $M_8$ as the reference one for the models $M_9$ and $M_{10}$.
The results are shown in Tab.~\ref{tab:limits_overlap_low}.

One can see that at least in the case of a simple power-law composition evolution,
lowering the energy threshold is beneficial both in the case of increasing and decreasing
proton fraction. In both cases lower energy bins do not spoil the separation achieved at higher energies, 
but can give a non-zero or even dominant contribution to the total separation power of the method.

\begin{table*}
\begin{center}
\begin{tabular}{|c|c|c|c|c|c|}
\hline
$E_{\rm min}$, EeV & $N_{\rm ev}$ & $P(M_5, M_6)$ & $P(M_5, M_7)$ & $P(M_8, M_9)$ & $P(M_8, M_{10})$ \\
\hline
57  & 140  & 0.13 & 0.57 & 0.09 & 1.00 \\ \hline
10  & 5000 & 0.17 & 0.94 & 0.10 & 1.00 \\ \hline
\end{tabular}
\end{center}
\caption{
Values of statistical power to constrain a test composition model at $95\%$~C.L. if data follows a reference model, see text for explanation.
Comparison of results for two different values of $E_{\rm min}$.
Binning over energy is 10 logarithmic bins per decade in both cases.
}
\label{tab:limits_overlap_low}
\end{table*}

\section{Conclusions}
\label{sec:discussion}

In summary, in this paper we proposed a quantitative method to assess composition of UHECR by using information on their arrival directions and energies, under the assumption that sources follow the large-scale matter distribution in the Universe. The key point of the proposal is calculation of the typical deflection angle with respect to the LSS source model. This angle is defined as a minimum $\theta_{\rm rec}$ of the likelihood function $TS(\theta)$, Eq.~(\ref{eq:TS}). It should be calculated for the data  and compared  to the same quantity calculated for the composition model in question. We have shown by applying
Eq.~(\ref{eq:TS}) to realistic mock event sets that the minimum is robust to the presence of the regular GMF and to mixed compositions, and therefore its position can be used to discriminate between composition models. 

To quantify the discriminating power of the test we calculated the standard statistical power for several pairs of the null-alternative models. We found that the statistical power reaches 1 (the maximum value) or gets close to 1 in a number of cases, in particular, when the alternative model is a pure proton composition. In other words, the distribution of $\theta_{\rm rec}$ for pure proton model is well separated from other models.
This means that the pure proton composition has good chances to be ruled out, in agreement with the results of Ref.~\cite{diMatteo:2017dtg}. In all cases we found that the discriminative power increases with the statistics of UHECR regardless of the assumed GMF model. 

Finally, we have investigated the dependence of the results on the unknown parameters, of which 
most important are those characterizing magnetic fields. We have seen that, while changing these parameters within reasonable limits does change the typical deflection angle, this change is not so large as to make it impossible to constrain the composition parameters or discriminate between models. In either case the conclusion strengthens with  the accumulation of statistics. 

Our method has several advantages: it is based exclusively on measured UHECR arrival directions and energies of events which are most reliably determined from the reconstruction of air showers; it is not sensitive to the details of the regular GMF  and to the presence of the non-extreme EGMF;  it can give conclusive results even at highest energies where use of other methods of composition study is limited by low UHECR statistics.

These advantages come at a price of having only one parameter determined from the data, so in general only one combination of variables characterizing composition can be determined/constrained. Note however that more parameters (in particular, those characterizing magnetic fields) can be incorporated into Eq.~(\ref{eq:TS}) in a straightforward way, trading model independence for additional information on composition.  Another way to increase the  constraining  power of the method is to simulate more realistic UHECR propagation taking into account secondary particles. We leave these interesting prospects for future study.

\acknowledgments
We would like to thank S.~Troitsky, O.~Kalashev and V.~Rubakov for useful discussions and helpful comments on the manuscript. This work is supported in part by the IISN, convention 4.4501.18
and in the framework of the State project ``Science'' by the Ministry of Science and Higher Education of the Russian Federation under the contract 075-15-2020-778.

\newpage
\bibliographystyle{JHEP}
\bibliography{ref}

\end{document}